\documentclass[12pt]{article}
\usepackage{latexsym,epsf}
\usepackage{amsmath}
\newenvironment{hangref}{\begin{list}{}{\setlength{\itemsep}{0pt}
\setlength{\parsep}{0pt}\setlength{\leftmargin}{+\parindent}
\setlength{\itemindent}{-\parindent}}}{\end{list}}
\begin{document}

\renewcommand{\baselinestretch}{2}
\normalsize

\title{An Epidemiological Approach to the \\Spread of Political Third Parties}
\author{Daniel M. Romero$^{1}$, Christopher M. Kribs-Zaleta$^{2*}$, \\
Anuj Mubayi$^2$, Clara Orbe$^3$ \\ 
\footnotesize $^{1}$ Center for Applied Mathematics,
Cornell University, Ithaca, NY \\
% dmr239@cornell.edu
\footnotesize $^{2}$ Department of Mathematics, 
University of Texas at Arlington, Arlington, TX\\
\footnotesize $^{3}$ Department of Applied Physics and Applied Mathematics, 
Columbia University, New York, NY \\
% co2203@columbia.edu
} \maketitle

* Corresponding author, e-mail: kribs@uta.edu

Mailing address: UTA Mathematics Department, Box 19408, Arlington, TX
76019-0408

Tel: +1 (817)272-5513, fax: +1 (817)272-5802

NOTE: E-mail is the preferred contact as Dr. Kribs-Zaleta will be working
at the Universit\'e de Lyon 1 in France during the 2009--2010 academic year.
The telephone and fax numbers in Lyon are not yet available.

\newpage
%%%%%%%%%%%%%%%%%%%%%%%%%%%%%%%%%%%%%%%%%%%%%%%%%%%%%%%%%%%%%%%%%%%%%%%%
%%                            ABSTRACT                                %%
%%%%%%%%%%%%%%%%%%%%%%%%%%%%%%%%%%%%%%%%%%%%%%%%%%%%%%%%%%%%%%%%%%%%%%%%

\begin{abstract}

Third political parties are influential in shaping American politics.  
In this work we study the spread of a third party ideology in a voting
population where we assume that party members/activists are more 
influential in recruiting new third party voters than non-member third 
party voters.  The study uses an epidemiological metaphor to develop
a theoretical model with nonlinear ordinary differential equations
as applied to a case study, the Green Party.  Considering long-term
behavior, we identify three threshold parameters in our model that describe
the different possible scenarios for the political party and its spread.
We also apply the model to the study of the Green Party's growth
using voting and registration data in six states and the District of Columbia
to identify and explain trends over the past decade.
Our system produces a backward bifurcation that helps identify conditions
under which a sufficiently dedicated activist core can enable a third party
to thrive, under conditions which would not normally allow it to arise.
Our results explain the critical role activists play in sustaining
grassroots movements under adverse conditions.  
\end{abstract}


\bigskip


\newpage
%%%%%%%%%%%%%%%%%%%%%%%%%%%%%%%%%%%%%%%%%%%%%%%%%%%%%%%%%%%%%%%%%%%%%%%%%
%%                         Section 1                                   %%
%%%%%%%%%%%%%%%%%%%%%%%%%%%%%%%%%%%%%%%%%%%%%%%%%%%%%%%%%%%%%%%%%%%%%%%%%

\section{Introduction}

% Why do we care?
The 2000 United States presidential election was for many a
testimony to the impact of third parties in a traditionally
bipartisan government.  Ralph Nader, the presidential candidate for
the Green Party, won 2\% of the popular vote, a percentage that many
attribute to the defeat of Democratic candidate Al Gore
%\cite{southwell2}
(Southwell, 2004).  The Green Party captured a seemingly
insignificant number of votes relative to majority percentages, yet
its presence in the election ultimately served to shape American
politics for the years following.  This incident demonstrates how
third parties, often emerging as grassroots movements (i.e.,
movements at the local level rather than at the center of major
political activity), can ultimately impact at the national level,
hence prompting the need to study their emergence and spread within
a voting population.

% What is a third party?
Third parties are defined as political parties operating along with
two major parties in a bipartisan system over a limited period of
time (where we define a limited period of time as a range of a few
years).  For the purposes of this paper we apply this definition to
all minor parties.  Traditionally, third parties have served as
venues of political dissent for voting individuals dissatisfied with
the major candidates in an election.  They often tackle specific
issues otherwise ignored by major political parties, thus
relinquishing popular support nationwide.  As Supreme Court Justice
Earl Warren wrote in 1957, ``History has amply proved the virtue of
political activity by minority, dissident groups, which innumerable
times have been in the vanguard of democratic thought and whose
programs were ultimately accepted'' 
%\cite{supreme}
(Warren, 1957).  Hence, while third
parties rarely capture the majority vote, their agendas, often
incorporated into major party platforms, are significant nonetheless.

% What are we doing?
Given the potential relevance of third parties to national politics,
we study, qualitatively and numerically, the dynamics of the
emergence and spread of third parties on a local level where growth
is measured in terms of the number of third party voters and members.  
We restrict our study to a local level because third parties usually 
originate in a small group and, via a ``bottom-up'' method of diffusion,
spread within a population by acquiring local official positions and 
then expanding to higher levels of government (Kowalewski, 1995).
%\cite{kowal}.  
Although individual personalities and circumstances dominate the initial
formation of any group, the ability of even a small group of people
to make itself heard within a larger community is great enough, especially
in an age of information technology, that we consider the local level
large enough to be described by a collective average.  That is, the
voting population under study is large enough that stochastic (random)
effects are dominated by the deterministic average behavior of the group.

% Explain epidemiological metaphor.
We use an epidemiological paradigm 
%\cite{brauer}
(Brauer and Castillo-Ch\'avez, 2001) to translate third party
emergence from a political phenomenon to a mathematical one where we assume
that third parties grow in a similar manner as epidemics in a population.  
We take this approach following in the steps of previous theoretical
studies that model social issues via such methods 
%\cite{bet,CCC,crisosto,Gonzalez,X}
(Bettencourt et al., 2006; Castillo-Ch\'avez and Song, 2003; 
Crisosto et al., 2001; Gonz\'alez et al., 2003; Song et al., 2006).
The epidemiological metaphor is suggested by the assumption that individuals'
decisions are influenced by the collective peer pressure generated by others'
behavior; the ``contacts'' between these two groups' ideas are analogous to
the contact processes that drive the spread of infectious diseases.
Here we assume that a certain subpopulation of the voting individuals, 
defined according to certain demographic factors, is more receptive
(in epidemiological terms, susceptible) to third party ideology than
the rest of the voting population, and that their political behavior 
is therefore driven by such collective peer-pressure contacts
%\cite{southwell1,timpone,wong}
(Southwell, 2003; Timpone, 1998; Wong, 2000).

% Daniel's additional explanation of how people make voting decisions
There are many components that may affect the decision of a person when it
comes to voting.  During a political campaign all candidates spend a great
amount of effort to assure that as many people as possible get exposed to their
ideas, and most importantly, their name.  People are often so overwhelmed by
exposure to different candidates that they end up depending more on
informal ways of obtaining information such as talking to other people.
This makes networking among people very influential during election time;
in fact some studies show that voting is ``contagious'' (Nickerson, 2008).
%\cite{dmr2}
It has been shown that people often rely on friends, relatives, coworkers,
etc. to obtain information about the candidates 
%\cite{dmr3}
(Robinson, 1976).  Also, there is
evidence that people who support a particular candidate tend to encourage
others to vote for that candidate 
%\cite{dmr1}
(Huckfeldt and Sprague, 1995). This makes the interaction
between people a very important factor in the voting process because both
the person wanting information seeks it from other people and the ones with
particular preferences want to share them.

%%% further background / lit. review
Collective behaviors such as voting have been studied for decades,
notably in the seminal work of Granovetter (1978),
%\cite{grano}, 
who considered
individuals in a population to have a distribution of thresholds with
regard to their willingness to participate in a particular collective
behavior (he used rioting as a primary example).  Other subsequent studies
considered the role played by a core group of especially influential
individuals (Macy, 1991; Oliver and Marwell, 1988), 
%\cite{macy,olmar}
such as party activists recruiting voters.
These studies used probabilistic and stochastic models which yielded
primarily numerical results; in applying the epidemiological framework
described above, we focus on the collective (rather than individual)
thresholds for persistence of a behavior (here voting and party membership)
which our deterministic models allow us to calculate.

% Explain application to Green Party.
While our model is designed to pertain to all third parties, we
consider the Green Party as a case study.  Although formally united
under the Association of State Green Parties in 1996 (and later the
nationalized Green Party of the United States in 2001), state-based
green parties have thrived in the U.S. at the local level since 1984,
when the Green Committees of Correspondence (CoC) were formed with the
purpose of organizing local Green groups and working toward the founding
of a national Green political party (Marks, 1997).
%\cite{marks}.
Our particular study focuses on the growth (and in some cases decline) of
the Green Party in six states and the District of Columbia in the past decade,
using voting and registration data.
In comparing the predictions of our model to particular data,
we consider a short time frame so that we can assume that social 
structure within the state in question does not change drastically, 
a necessary condition for assuming voting population heterogeneity.

% Map out the article.
We organize our paper as follows:  Section~2 describes our theoretical
model, as well as a simplification in which the entire population
under study is equally receptive to third party ideas.
Section~3 presents the mathematical analysis of the simplified model
(which employs qualitative analysis techniques from the field of
nonlinear dynamical systems, as well as sensitivity analysis)
and interprets the results, and Section~4 applies them to a case study,
using data to estimate model parameters.  Section~5 performs a similar
but more limited analysis of the more complex model, and Section~6
draws conclusions about the implications of our models for the growth
of third parties from grassroots movements.

%%%%%%%%%%%%%%%%%%%%%%%%%%%%%%%%%%%%%%%%%%%%%%%%%%%%%%%%%%%%%%%%%%%%%%%%
%%             Section 2                                              %%
%%%%%%%%%%%%%%%%%%%%%%%%%%%%%%%%%%%%%%%%%%%%%%%%%%%%%%%%%%%%%%%%%%%%%%%%

\section{A Population Model for the Spread of a Third Party}

\subsection{Underlying Assumptions}\label{assume}

In developing our model, we apply epidemiological terminology to describe 
the growth of a third party.  The assumptions we make about how individuals
behave, and change their behavior, define the classes in our compartmental
model and the rates at which individuals move between classes.

\begin{enumerate}
\item[(1)] We assume that our population is a heterogeneous mix of individuals 
who belong to different backgrounds according to certain demographic factors.
\end{enumerate}

% population divided into two susceptibility levels; what the levels are
Our model considers a population of all voters, $N$, divided into
two classes or sub-populations whose susceptibility to third party
ideology is based on demographic factors such as education,
socioeconomic status, race, gender, age, political orientation and
professional occupation.  Inherently, certain demographic characteristics,
labelled as high affinity, make an individual more likely to subscribe 
to a third party's ideology, which targets a more specific audience
than alternative majority agendas.  
That is, upon entering the voting system, certain individuals are
more statistically inclined to vote a certain way.  For example, 
a progressive environmental activist is statistically more likely 
to agree and vote for the Green Party agenda, which stresses 
communal-based economics, local government, and gender and racial equity, 
than a conservative corporate executive whose economic philosophy 
directly conflicts with that of the Green Party.
For this reason we consider population heterogeneity vital to this study.  

We apply the following method of dividing the entering voting
population into two susceptible classes: if an individual has more
high affinity factors than low affinity factors then that person
directly enters the high affinity class and similarly for low
affinity susceptibles.  We define high affinity factors as features
of the individual based on his/her demographic profile that make
him/her more inclined to vote for the third party; conversely, low
affinity factors make the individual less statistically likely to
subscribe to the party's platform.  We assume that a constant proportion
$p$ ($0<p<1$) of new voters enters the high-affinity class $H$
(the remaining $1-p$ proportion enter $L$).

\begin{enumerate}
\item[(2)] We assume that individuals' affinity factors (demographic
  characteristics) remain fixed for the period of the study.
\item[(3)] We assume that a third party's agenda remains consistent over time.
\end{enumerate}

% affinity level fixed
We assume that individuals do not move from one susceptibility class
to the other.  One reason for this is the relative permanence of
individuals' demographic characteristics.  We limit our model to tracing the
{\em expansion} of the third party; hence, we refer to a shorter time period
over which we assume social structure remains constant.  In other words,
individuals with high affinity to the third party do not become individuals
with low affinity, and vice versa.

% Third party platforms
The other reason for this consistency has to do with the parties themselves.
In addition to being more specific than major party agendas 
(i.e., more specific in their goals and less geared to moderacy), 
third party platforms tend to be more consistent over time.  
Third parties are not pressured to constantly adjust to the shifting 
demands of the populace since they do not seek the majority vote.  
Consequently, they do not target the majority voting population.
Each party has its own agenda, which appeals to certain sectors of 
the voting population.  Hence different parties target voting populations 
that are more inclined to subscribe to their ideology.  While one party, 
for example, may target individuals from a certain educational background 
that we, in our model, label as high affinity and that other parties may
overlook, all parties nonetheless recognize that education factors into 
an individual's likelihood to support or refute that party's platform.  
It is true that individuals from varied backgrounds comprise the main
parties, yet, when dealing with the specific agendas of third parties that 
do not strive to sway the majority vote, we assume that third parties 
appeal to individuals of certain demographic backgrounds more than others.
Therefore, we account for the aforementioned standard set of
demographic factors that parties look at when spreading their
ideologies.  In our paper we apply our model to an individual case
study of the Green Party of Pennsylvania; however, the same methodology
of distinguishing susceptibles can be applied to all third parties.

% define party
\begin{enumerate}
\item[(4)] We define two levels of participation in third-party politics:
voting for third-party candidates, and membership.
A party exists only if it has members; we define members as those 
who pay dues, volunteer, and preside over party affairs.
\end{enumerate}

% infected classes
As described above, all voters enter the voting system either to the 
low affinity, $L$, or high affinity, $H$, susceptible class.  According
to our epidemiological metaphor, in addition to these two susceptible classes,
our model includes three infected classes: $V_H$, $V_L$, and $M$, 
third party voters from the high affinity class, third party voters 
from the low affinity class, and party members respectively.  We define 
party members as voters of the third party who pay dues to the party;
often such members officiate, volunteer and actively campaign for
voter recruitment.  In epidemiological terms, $V_H$ and $V_L$ 
correspond to voters of a lower degree of infection and individuals 
of the $M$ class are voters infected to a higher degree.  We distinguish
between $V_H$ and $V_L$ because of their interactions with their
respective ``neighbors'' in $H$ and $L$.

\begin{enumerate}
\item[(5)] We assume that third parties, emerging through resource-limited
grassroots efforts, spread primarily via primary (direct) contacts between
third-party supporters and susceptibles.
%\item[(6)] We assume that the influence of third party voters on susceptibles
%who become third party voters is the same regardless of their respective (high
%or low) affinity classes.
%However, since party members correspond to the higher degree of
%`infection,' members have a greater effect in voter recruitment.
\item[(6)] We assume that third party members have a greater effect upon
voter recruitment than do third party voters, due to members' activism.
\end{enumerate}

% infection transitions
In our model, a system of nonlinear differential equations,
we consider susceptible movement into voting and member compartments 
as well as possible regressions back from the third party voting phase 
into the susceptible class.  Once an individual is susceptible
he/she can become `infected' (either $V_H$ or $V_L$) through direct
contact with the $V_H$, $V_L$, and $M$ classes.  
Due to a lack of funding and resulting lack of mass-media exposure 
to the general population, these primary contacts with susceptibles
involve such direct interaction as personal meetings, phone conversations, 
and electronic communications like personally addressed e-mails
and weblog (blog) comments.  We assume that the rate at which these
contacts occur is proportional both to the size of the susceptible group
and to the proportion of third-party supporters in the population.

We do not consider a linear term weighing the influence of media coverage 
from the third party (i.e., secondary contact factors) in the forward
transition from both susceptible classes to third party voting classes.
Instead, we focus on the nonlinear terms considering
the effects of voters from the $V_H$, $V_L$, and $M$ classes, where
voters from $V_H$ and $V_L$ bear an 
%equal 
affinity-specific influence $\beta_H$ (from $H$ to $V_H$) 
and $\beta_L$ (from $L$ to $V_L$) in third party voter
recruitment.  Through activism, members from $M$ influence susceptibles 
of each type at higher rates ($\alpha \beta_H$ and $\alpha \beta_L$,
respectively) than voters, with their increased influence measured by
the parameter $\alpha$ ($\alpha>1$).

\begin{enumerate}
\item[(7)] We consider both primary (direct) contacts and secondary (indirect)
contacts in the regression of third party voters to the susceptible class.
\item[(8)] We assume that all other parties exert equal influence in
discouraging third party voting.
%\item[(9)] We assume that individuals from the same susceptibility class
%have more influence in discouraging third party voters of each background
%than individuals from the other susceptibility class.
\item[(9)] We assume that individuals have more influence upon others of
their same affinity class (high or low) than upon members of the other
affinity class, in encouraging and discouraging third party voting.
\end{enumerate}

% defection transitions
We consider the transitions back from the third party voting to the
susceptible classes to involve both linear terms, $\epsilon_H V_H$
and $\epsilon_L V_L$, and nonlinear terms, $\phi_H(H + \sigma L)
\frac{V_H}{N}$ and $\phi_L(L +\sigma H)\frac{V_L}{N}$,
contributions by secondary contacts with the opposition (i.e., media
from well-funded majority voters) and direct contact with the
susceptible classes respectively.
Compared to primary contacts, described in a previous paragraph, secondary
contacts include mass e-mails, media, and circulating literature.

Voters from a certain susceptibility class (with its own set of 
demographic factors) address issues that usually appeal more to 
%third party 
voters deriving from the same class.  Therefore,
susceptible individuals with high affinity bear a greater influence in
recruiting voters who came from the high affinity susceptible class back 
into the susceptible class than susceptible individuals with low affinity.
%Similar to $\alpha$ in the forward transition, we designate $\tau$ as the
%augmentation factor.  Therefore, in regressing back from $V_H$ to
%$H$, $\tau H$ represents the greater influence that $H$ individuals
%exert on voters from $V_H$ than susceptibles from $L$, the lower affinity 
The reduction in influence by individuals from a different affinity class
is denoted by $\sigma$, so that in regressing back from $V_H$ to $H$,
$\sigma L$ represents the lesser influence that $L$ individuals exert
on voters from $V_H$ than do susceptibles from $H$, the higher affinity
class (similar reasoning applies to the $V_L$-to-$L$ transition).
Likewise, the cross-affinity influence in recruiting third party voters
is reduced by a factor of $\sigma$.

\begin{enumerate}
\item[(10)] We assume that third party voters become active party members
through the ongoing efforts (primary contacts) of the members, 
who have made a permanent commitment to the party.
\end{enumerate}

% V to M transitions
Once voting for the third party, individuals can become party members.  
They enter this higher state of infection via the nonlinear terms 
$\gamma V_H \frac{M}{N}$ and $\gamma V_L \frac{M}{N}$, where
we only consider the influence of primary contacts with party members
in bringing about this transition, measured by the rate parameter $\gamma$.  
Given that we are studying the spread of the party, we assume that 
party members do not resign their memberships.  We reason that 
once an individual feels strongly enough to join a party, he/she retains 
his/her loyalty to the party; the only way a person stops being a member 
(during the growth period under study) is by leaving the voting system.

% demographic renewal
Finally, we consider natural exits from all classes as a result of
death or moving.  The sum of the equations of the model, for both
versions developed below, gives $\frac{dN}{dt}=0$, verifying that
our population stays constant, a safe assumption by which the number
of people entering the voting system (i.e., coming of age, moving
in) counterbalances the number of people leaving the system (i.e.,
dying, moving out).

\subsection{The General (Two-Track) Model}

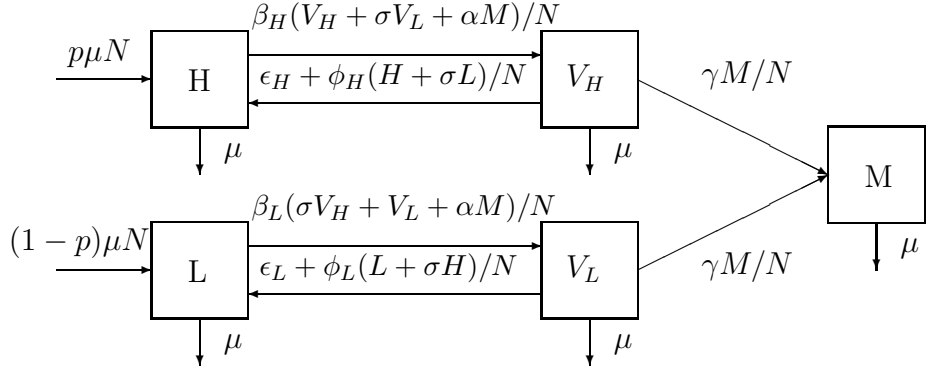
\begin{figure}[th]
\begin{centering}
%\leavevmode
%\includegraphics[width = 6 in, height = 3.5 in]{model.jpg}
\setlength{\unitlength}{3947sp}%
\begin{picture}(5767,2259)(-354,-2473)
%\begin{picture}(width,height)(LLx,LLy)
\thinlines
% entering H
\put( -49, -661){\vector( 1, 0){600}}
\put(  26, -586){\makebox(0,0)[lb]{$p\mu N$}}
% entering L
\put( -49,-1861){\vector( 1, 0){600}}
\put(-349,-1786){\makebox(0,0)[lb]{$(1-p)\mu N$}}
% H box with mortality
\put( 551, -961){\framebox(600,600){}}
\put( 776, -736){\makebox(0,0)[lb]{H}}
\put( 851, -961){\vector( 0,-1){300}}
\put(1001,-1186){\makebox(0,0)[lb]{$\mu$}}
% L box with mortality
\put( 551,-2161){\framebox(600,600){}}
\put( 776,-1936){\makebox(0,0)[lb]{L}}
\put( 851,-2161){\vector( 0,-1){300}}
\put(1001,-2386){\makebox(0,0)[lb]{$\mu$}}
% H-VH flows
\put(1151, -511){\vector( 1, 0){1850}}
\put(1176, -361){\makebox(0,0)[lb]{\small $\beta_H(V_H+\sigma V_L+\alpha M)/N$}}
\put(3001, -811){\vector(-1, 0){1850}}
\put(1226, -736){\makebox(0,0)[lb]{\small $\epsilon_H+\phi_H(H+\sigma L)/N$}}
% L-VL flows
\put(1151,-1711){\vector( 1, 0){1850}}
\put(1176,-1561){\makebox(0,0)[lb]{\small $\beta_L(\sigma V_H+V_L+\alpha M)/N$}}
\put(3001,-2011){\vector(-1, 0){1850}}
\put(1226,-1936){\makebox(0,0)[lb]{\small $\epsilon_L+\phi_L(L+\sigma H)/N$}}
% VH box with mortality
\put(3001, -961){\framebox(600,600){}}
\put(3151, -736){\makebox(0,0)[lb]{$V_H$}}
\put(3301, -961){\vector( 0,-1){300}}
\put(3451,-1186){\makebox(0,0)[lb]{$\mu$}}
% VL box with mortality
\put(3001,-2161){\framebox(600,600){}}
\put(3151,-1936){\makebox(0,0)[lb]{$V_L$}}
\put(3301,-2161){\vector( 0,-1){300}}
\put(3451,-2386){\makebox(0,0)[lb]{$\mu$}}
% V-M flows
\put(3601, -661){\vector( 2,-1){1200}}
\put(4001, -736){\makebox(0,0)[lb]{$\gamma M/N$}}
\put(3601,-1861){\vector( 2, 1){1200}}
\put(4001,-1936){\makebox(0,0)[lb]{$\gamma M/N$}}
% M box with mortality
\put(4801,-1561){\framebox(600,600){}}
\put(5026,-1336){\makebox(0,0)[lb]{M}}
\put(5101,-1561){\vector( 0,-1){300}}
\put(5251,-1786){\makebox(0,0)[lb]{$\mu$}}
\end{picture}%
\caption{The two-track model, with per capita flow rates} \label{model}
\end{centering}
\end{figure}

We first introduce a two-track model, as described immediately above, 
to study the dynamics between a heterogeneously mixed population of 
susceptible voters, third party voters, and party members.
We apply the following set of ordinary differential equations to
model voting dynamics, as illustrated in Figure~\ref{model}.

\begin{align}
\frac{dH}{dt}&=p\mu N+\epsilon_H V_H+\phi_H(H+\sigma L)\frac{V_H}{N}
-\beta_H(V_H+\sigma V_L+\alpha M)\frac{H}{N}-\mu H, \\
\frac{dL}{dt}&=(1-p)\mu N+\epsilon_L V_L+\phi_L(L+\sigma H)\frac{V_L}{N}
-\beta_L(\sigma V_H+V_L+\alpha M)\frac{L}{N}-\mu L, \\
\frac{dV_H}{dt}&=\beta_H(V_H+\sigma V_L+\alpha M)\frac{H}{N}-\epsilon_H
V_H-\phi_H(H+\sigma L)\frac{V_H}{N}-\frac{\gamma M V_H}{N}-\mu V_H, \\
\frac{dV_L}{dt}&=\beta_L(\sigma V_H+V_L+\alpha M)\frac{L}{N}-\epsilon_L
V_L-\phi_L(L+\sigma H)\frac{V_L}{N}-\frac{\gamma M V_L}{N}-\mu V_L, \\
\frac{dM}{dt}&=\frac{\gamma M V_H}{N}+\frac{\gamma M V_L}{N}-\mu M,
\label{meqn0} \\
N&=H+L+V_H+V_L+M.
\end{align}
Adding equations (1), (2), (3), (4) and (5) yields $\frac{dN}{dt}=0$, 
showing that the total population $N$ is constant over time.
Model parameters are summarized in Table~1.
%\ref{p2t}. %% hard-coded to 1 because crazy JMP ordering of tables after
%appendix screws up LaTeX's ability to cite the table correctly.  Why?????

\subsection{The Simplified (One-Track) Model}
In order to facilitate analysis of the two-track model, we initially
consider a simplified version that does away with voting population 
heterogeneity (assumption (1), and consequently (9)) before exploring 
analysis for the more complex system.  This simplified model assumes 
a homogeneous susceptible population ($p=0$ or $p=1$), reducing 
the two-track model to one susceptible class, $S$, and two infected classes: 
third party voters, $V$, and party members, $M$, respectively.  The $S$ class 
comprises those individuals who vote, but do not vote for the third party.  
The $V$ class comprises the third party voters, and the $M$ class again has 
third party members (i.e., party officials, donors, volunteers).  

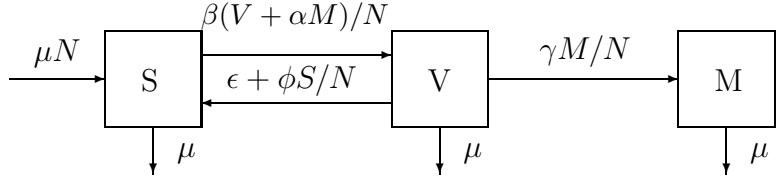
\begin{figure}[th]
\begin{centering}
%\leavevmode
%\includegraphics[width = 6 in, height = 1.5 in]{simp-model.jpg}
\setlength{\unitlength}{3947sp}%
\begin{picture}(4824,1062)(-11,-1276)
\thinlines
% incoming
\put(   1,-661){\vector( 1, 0){600}}
\put( 151,-586){\makebox(0,0)[lb]{$\mu N$}}
% S box with mortality
\put( 601,-961){\framebox(600,600){}}
\put( 826,-736){\makebox(0,0)[lb]{S}}
\put( 901,-961){\vector( 0,-1){300}}
\put(1051,-1186){\makebox(0,0)[lb]{$\mu$}}
% S-V flows
\put(1201,-511){\vector( 1, 0){1200}}
\put(1201,-361){\makebox(0,0)[lb]{\small $\beta(V+\alpha M)/N$}}
\put(2401,-811){\vector(-1, 0){1200}}
\put(1361,-766){\makebox(0,0)[lb]{$\epsilon+\phi S/N$}}
% V box with mortality
\put(2401,-961){\framebox(600,600){}}
\put(2626,-736){\makebox(0,0)[lb]{V}}
\put(2701,-961){\vector( 0,-1){300}}
\put(2851,-1186){\makebox(0,0)[lb]{$\mu$}}
% V-M flow
\put(3001,-661){\vector( 1, 0){1200}}
\put(3326,-586){\makebox(0,0)[lb]{$\gamma M/N$}}
% M box with mortality
\put(4201,-961){\framebox(600,600){}}
\put(4426,-736){\makebox(0,0)[lb]{M}}
\put(4501,-961){\vector( 0,-1){300}}
\put(4651,-1186){\makebox(0,0)[lb]{$\mu$}}
\end{picture}
\caption{The one-track model, with per capita flow rates} \label{simp-model}
\end{centering}
\end{figure}

In the one-track model we omit unnecessary parameters from the heterogeneous 
version.  Figure~\ref{simp-model} illustrates the one-track model, and
Table~\ref{p1t} summarizes the parameters.

%%% table 2 was here

In this case the model reduces to the following system, which{\small\ }is
effectively{\small\ }two-dimensional since $N$ can again be seen 
to be constant:
\begin{align}
\frac{dS}{dt}&=\mu N+\epsilon V+\phi S\frac{V}{N}
-\beta(V+\alpha M)\frac{S}{N}-\mu S, \label{S1}\\
\frac{dV}{dt}&=\beta(V+\alpha M)\frac{S}{N}
-\epsilon V-\phi S\frac{V}{N}-\frac{\gamma M V}{N}-\mu V, \label{V1}\\
\frac{dM}{dt}&=\frac{\gamma M V}{N}-\mu M,\label{M1}\\
N&=S+V+M.
\end{align}
%Adding equations (7), (8), and (9) yields $\frac{dN}{dt}=0$, 
%showing that the total population $N$ is constant over time.

%%%%%%%%%%%%%%%%%%%%%%%%%%%%%%%%%%%%%%%%%%%%%%%%%%%%%%%%%%%%%%%%%%%%%%%%
%%             Section 3                                              %%
%%%%%%%%%%%%%%%%%%%%%%%%%%%%%%%%%%%%%%%%%%%%%%%%%%%%%%%%%%%%%%%%%%%%%%%%

\section{Analysis of the One-Track Model}\label{sec3}

We begin our analysis by calculating equilibria for our model
and determining conditions for their existence and stability.
We first simplify the system in two ways.  Since the total voting
population $N$ is assumed constant, we can reduce our system 
to two dimensions by rewriting $S=N-V-M$ in equation \eqref{V1}, so that
\begin{equation}
\frac{dV}{dt}=\beta(V+\alpha M)\frac{N-V-M}{N}
-\epsilon V-\phi(N-V-M)\frac{V}{N}-\frac{\gamma M V}{N}-\mu V.\label{V1a}
\end{equation}
We can now analyze the system defined by \eqref{V1a} and \eqref{M1},
as $S$ can always be found once $V$ and $M$ are known.

The constancy of $N$ also allows us to proportionalize the
system by defining new variables $v=\frac{V}{N}$, $m=\frac{M}{N}$,
and $s=\frac{S}{N}=1-v-m$, which give the proportion of the population
in each class.  Dividing equations \eqref{V1a} and \eqref{M1} by $N$
and substituting the new variables gives, finally, the system
\begin{align}
\frac{dv}{dt}&=\beta(v+\alpha m)(1-v-m)
-\epsilon v-\phi(1-v-m)v-\gamma m v-\mu v,\label{v1b}\\
\frac{dm}{dt}&=(\gamma v-\mu)m.\label{m1b}
\end{align}
In order to analyze stability we linearize the system and compute
partial first derivatives with respect to each of the variables, $v$ and $m$,
obtaining the Jacobian matrix $J_1$ for system \eqref{v1b}--\eqref{m1b}:
$$J_1=
\begin{pmatrix}(\beta-\phi)(1-2v-m)-(\alpha\beta+\gamma)m-(\mu+\epsilon)
&\hspace{10 pt}\alpha\beta(1-v-2m)-(\beta-\phi+\gamma)v\\
& \\
\gamma m & \gamma v-\mu\\
\end{pmatrix}.$$

\subsection{$E_1$: Party-Free Equilibrium (PFE)}
The party-free equilibrium (PFE) for the reduced system
occurs at (0,0), the steady state achieved when the entire
population resides in the $S$ class (i.e., the third party has
neither voters nor members and, by definition of party existence,
does not exist).  The PFE is essentially analogous to the
disease-free equilibrium in epidemiology and always exists 
as a possible outcome for the voting population.

Applying the above reduced Jacobian matrix to our PFE, (0,0), where
we only consider the $v$ and $m$ terms, we determine PFE stability:
$$J_1(0,0)=
\begin{pmatrix}
(\beta-\phi)-(\mu+\epsilon)&\alpha\beta\\
& \\
0& -\mu
\end{pmatrix}.
$$
The equilibrium point (0,0) will be locally asymptotically stable
(LAS) if all the eigenvalues of the matrix are negative.  Assuming
$\mu>0$, the eigenvalue $-\mu$ of the Jacobian is always negative,
whereas the second eigenvalue $(\beta-\phi)-(\mu+\epsilon)<0$ if and
only if $\frac{(\beta-\phi)}{\mu+\epsilon}<1$.  For ease of notation
and interpretation we define the threshold quantity 
$R_1=\frac{(\beta-\phi)}{\mu+\epsilon}$, so that the PFE is LAS
if and only if $R_1<1$.  We discuss the relevance of this threshold value 
in the last part of this section.

\subsection{$E_2$: Member-Free Equilibrium (MFE)}
The member-free equilibrium (MFE) occurs when $M=0$ but $V,S \neq 0$, 
i.e., the voting population subdivides between susceptibles, $S$, and 
third party voters, $V$.  While mathematically possible, this outcome is
politically unrealistic given that voters cannot vote for a party
that does not exist (recall our assumption (4) that party existence 
depends on the presence of an $M$ class).  For mathematical consistency, 
however, we consider the equilibrium point ($v_2^*$,$m_2^*$), 
where {$m_2^*=0$}.  (Note that the asterisk superscript denotes equilibrium
values, while the numerical subscript distinguishes the equilibrium point
in question.)  This arises from the equilibrium condition obtained 
by setting $dm/dt=0$ in \eqref{m1b}:
\begin{equation}
(\gamma v^*-\mu)m^*=0,\label{eqcondM}
\end{equation}
which implies that either $(\gamma v^*-\mu)=0$ (which we will consider later)
or $m^*=0$.

To find $v^*_2$ we set $dv/dt=0$ and $m^*=0$ in \eqref{v1b} and
rearrange terms to get
$$(\beta-\phi)v^{\ast 2}+(\mu +\epsilon +\phi -\beta)v^*=0.$$
This implies that either $v^*=0$ (the party-free equilibrium) or
$(\beta -\phi)v^* =\beta -(\mu +\epsilon +\phi)$.
We consider the situation where $v^*\neq 0$, solve for $v^*$ and
simplify the results as follows:
$$v_2^*= 1- \frac{\mu + \epsilon}{\beta -\phi}=1-\frac{1}{R_1}$$
where $v_2^*$ retains political value only if $R_1>1$---otherwise
$v_2^*<0$ which is meaningless.

Finally, we can also write
%solve for $s^*_2$ and obtain
$s^*_2=1-v^*_2-m^*_2=\frac{\mu+\epsilon}{\beta-\phi}=\frac{1}{R_1}$, 
which makes sense politically only if $R_1>1$ (i.e, $s_2^*<1$).  
We can therefore express our member-free equilibrium as 
$E_2=(1-\frac{1}{R_1},0)$.
The MFE exists if and only if $R_1>1$, since ignoring this condition 
leads to a negative third party voting population.
%If $R_1>1$, then $\frac{1}{R_1}<1$ (i.e., the
%entire population does not reside in the susceptible class).  This
%implies that the population has moved out of $S$ into the $V$ and
%$M$ classes.  Since $M=0$, the left over proportion (i.e.,
%$1-\frac{1}{R_1}$) resides in $V$.

The above situation makes mathematical sense but not political sense
since parties, by our original assumption, do not exist without
members, and in this member-free case we deal with voters who vote
for a non-existent party.  We might interpret this situation as having
voters still willing to vote for this party, but no party candidates
for whom to vote.

Regardless of the political likelihood of MFE existence, we consider
its stability.  Again, we apply the method of using the reduced
system's Jacobian matrix in determining the stability of the
member-free equilibrium:
$${J_1({\textstyle 1-\frac{1}{R_1}},0)=
\begin{pmatrix}
(\mu+\epsilon)(1-R_1)& &
\alpha\beta(\frac{1}{R_1})-(\beta-\phi+\gamma)(1-\frac{1}{R_1})\\
 & & \\
0& & \gamma(1-\frac{1}{R_1})-\mu\\
\end{pmatrix}.
}$$
The reduced system equilibrium point $(1-\frac{1}{R_1},0)$ is
locally asymptotically stable if all the eigenvalues of the above
matrix are negative.  We know that, since $R_1>1$ (in order for the
MFE to exist), one of the eigenvalues, $(\mu+\epsilon)(1-R_1)$, is
negative.  The second eigenvalue of the Jacobian is
$\gamma(1-\frac{1}{R_1})-\mu$.  This eigenvalue is negative if and
only if $\frac{\gamma}{\mu}(1-\frac{1}{R_1})<1$.

We define the left hand side of the inequality as
$R_2=\frac{\gamma}{\mu}(1-\frac{\mu+\epsilon}{\beta-\phi})
    =\frac{\gamma}{\mu}(1-\frac{1}{R_1})$.
Hence we have derived two threshold parameters $R_1$ and $R_2$ that
determine equilibria stability depending on relative parameter values.
(Note that they are related: $R_1=1\Leftrightarrow R_2=0$.)

\subsection{$E_3$ and $E_4$: Survival Equilibria}\label{E3E4sec}
In the event of survival equilibria, the voting population
subdivides between susceptibles, $S$, third party voters, $V$, and
members, $M$.  We regard this as a successful state of coexistence
and, given certain conditions, the point at which the party thrives.
We determine the equilibrium proportions by returning to the condition
\eqref{eqcondM} that $dm/dt=0$.  Since $\frac{dm}{dt}=(\gamma v^*-\mu)m^*$,
$v^*=\frac{\mu}{\gamma}$ when $m^*\neq0$.  Here we impose the condition
$\mu<\gamma$ so that $v^*<1$, since $s^*+v^*+m^*=1$.

Next we set $dv/dt=0$ and $v^*=\frac{\mu}{\gamma}$ in \eqref{v1b}:
% and rearrange terms to solve for $m^*$:
$$\frac{dv}{dt}=
(\beta-\phi)\,\frac{\mu}{\gamma}\,\left(1-\frac{\mu}{\gamma}-m^*\right)
+\alpha\beta m^*\left(1-\frac{\mu}{\gamma}-m^*\right)
-(\mu+\epsilon+\gamma m^*)\,\frac{\mu}{\gamma}=0.$$
This equation, which is quadratic in $m^*$, can be rewritten (after dividing
through by $-\alpha\beta$) in the form $f(m^*)=m^{*2}+BM^*+C=0$, where
\begin{equation}
B=\left(\frac{\beta-\phi}{\alpha\beta}\right)\frac{\mu}{\gamma}     
-\left(1-\frac{\mu}{\gamma}\right)+\frac{\mu}{\alpha\beta}, \;\; 
C=-\frac{\mu}{\gamma}\left[\frac{\beta-\phi}{\alpha\beta}
\left(1-\frac{\mu}{\gamma}\right)-\frac{\mu+\epsilon}{\alpha\beta}\right].
\label{BandC}
\end{equation}
Solutions to $f(m^*)=0$ are given by the quadratic formula
$$m^*_\pm=\frac{1}{2}\left[-B\pm\sqrt{B^2-4C}\right];$$
however, in general, these solutions may be real, or may fall outside
the meaningful interval $(0,1-v^*]$.  Depending on the values of model
parameters, there may be 0, 1, or 2 solutions within this interval,
each corresponding to a meaningful equilibrium with positive party membership.

Analysis of the conditions involved in determining the number of survival
equilibria is considerably more involved than that for $E_1$ and $E_2$;
details are given in the Appendix.  The results, which introduce an additional
threshold quantity, can be summarized as follows.

\bigskip

\noindent{\bf Proposition 1.}  {\em (i) If $R_2>1$, the system
  \eqref{v1b}--\eqref{m1b} has precisely one survival equilibrium
  $E_3=(\frac{\gamma}{\mu}, m^*_+)$.
%1-\frac{\gamma}{\mu}-m^*_+, 

(ii) If $R_2<1$, then the system has two survival equilibria, $E_3$ and
  $E_4=(\frac{\gamma}{\mu}, m^*_-)$, if and only if $R_3>1$, where
%1-\frac{\gamma}{\mu}-m^*_-,
%$B<0$ and $B^2-4C\ge 0$.  
\begin{equation}
R_3=\min\left(R_{3a},R_{3b}\right), \;\;
R_{3a}=r_3\left(1-\frac{1+q}{r_2}\right), \;\;
R_{3b}=\sqrt{r_3}\left(1-\sqrt{\frac{1-q}{r_2}+h}\,\right),
\label{r3def}
\end{equation}
and
$$q=\frac{\beta-\phi}{\alpha\beta}, \;\; 
  r_2=\gamma/\mu, \;\; r_3=\alpha\beta/\mu,\;\;
  h=\frac{2}{\sqrt{r_3}}\left(\sqrt{1+\frac{\epsilon}{\gamma}}-1\right).$$
  Otherwise there are none.}

\bigskip

To interpret the conditions $R_{3a}>1$ and $R_{3b}>1$, 
we can rewrite them as follows:
\begin{align}
B<0\Leftrightarrow R_{3a}>1\Leftrightarrow & \frac{1}{\alpha\beta}
+\frac{1+q}{\gamma}<\frac{1}{\mu}\,,
\label{Bcond-main}\\
B^2-4C\ge 0\Leftrightarrow R_{3b}>1\Leftrightarrow & 
\sqrt{\frac{1}{\alpha\beta}}+\sqrt{\frac{1-q}{\gamma}+\hat{h}(\epsilon/\gamma)}
\le\sqrt{\frac{1}{\mu}}\,,
\label{discrimcond-main}
\end{align}
where
$$\hat{h}(\epsilon/\gamma)=2\sqrt{\frac{1}{\alpha\beta}}\sqrt{\frac{1}{\mu}}
\left(\sqrt{1+\frac{\epsilon}{\gamma}}-1\right)>0.$$

%Note that the conditions in (ii) require $\frac{\alpha\beta}{\mu}>\max\left(
%1,\frac{\beta-\phi}{\mu+\epsilon}\,\frac{\beta-\phi-\epsilon}{\mu}\right)$.

Both inequalities \eqref{Bcond-main} and \eqref{discrimcond-main} relate 
the minimum amounts of time taken for an individual to be influenced 
by a party member ($M$) to move from $S$ to $V$, $1/\alpha\beta$, 
and from $V$ to $M$, $1/\gamma$, to the average lifetime of an individual 
in the voting system, $1/\mu$.  In order for the party to survive, 
a weighted sum of the first two times must be less than the average 
lifetime in the system, or else individuals in $S$ will, on average, 
leave the system before they can ``replace'' the members in $M$ who 
recruited them.  The weights in the sum involve 
$q=\frac{\beta-\phi}{\alpha\beta}$, the relative effectiveness of voters $V$
to members $M$ in recruiting new voters from $S$, because the influence of
party voters reduces the recruiting threshold burden on party members $M$
to some extent.\footnote{
If we define $T_1=1/\alpha\beta$, $T_2=1/\gamma$, and $T_3=1/\mu$,
\eqref{Bcond-main} and \eqref{discrimcond-main} can be written more simply
as $T_1+(1+q)T_2<T_3$, $\sqrt{T_1}+\sqrt{(1-q)T_2+h}\le\sqrt{T_3}.$}
(Note that $0<q<1$.)

Observe that \eqref{discrimcond-main} implies that 
\begin{equation}
\sqrt{\frac{1}{\alpha\beta}}+\sqrt{\frac{1-q}{\gamma}}\le\sqrt{\frac{1}{\mu}}.
\label{discrimcond0}
\end{equation}
If $\epsilon<<\gamma$, we can generate a Maclaurin (Taylor) expansion in
$\epsilon/\gamma$ for \eqref{discrimcond-main} which yields
$$\sqrt{\frac{1}{\alpha\beta}}+\sqrt{\frac{1-q}{\gamma}}
+\frac{1}{2}\,\frac{\sqrt{\frac{1}{\alpha\beta}}\sqrt{\frac{1}{\mu}}}
{\sqrt{\frac{1-q}{\gamma}}}\,
\left(\frac{\epsilon}{\gamma}\right)
+{\cal O}\left[\left(\frac{\epsilon}{\gamma}\right)^2\right]
\le\sqrt{\frac{1}{\mu}}.$$
This expansion suggests more quantitatively how the influence $\epsilon$
of voters for other parties $S$ in getting third-party voters $V$ to ``defect''
back to the major parties complicates the third party's survival (which
would otherwise only require \eqref{discrimcond0}).

\bigskip

Having established conditions for the existence of the two survival
equilibria, $E_3$ and $E_4$,
%$E_3=(1-\frac{\gamma}{\mu}-m^*_+, \frac{\gamma}{\mu}, m^*_+)$ 
%$E_4=(1-\frac{\gamma}{\mu}-m^*_-, \frac{\gamma}{\mu}, m^*_-)$, 
we examine their stability.  The reduced Jacobian matrix for the survival
equilibria follows:
$$J_1({\textstyle\frac{\mu}{\gamma}},m^*_\pm)=
\begin{pmatrix}
\left[(\beta-\phi)\left(1-2\frac{\mu}{\gamma}-m^*_\pm\right)\right.
\hspace*{10pt}&
\left[\alpha\beta\left(1-\frac{\mu}{\gamma}-2m^*_\pm\right)\right.\\
\left.-(\alpha\beta+\gamma)m^*_\pm-(\mu+\epsilon)\right]&
\left.-(\beta-\phi)\frac{\mu}{\gamma}-\mu\right]\\
& \\
\gamma m^*_\pm & 0\\
\end{pmatrix}.$$
The stability (LAS) criterion that the eigenvalues of this matrix have
negative real part is equivalent to the conditions that the trace be negative
and the determinant positive.  We calculate
$$\det J_1({\textstyle\frac{\mu}{\gamma}},m^*_\pm)=-\gamma m^*_\pm
  \left[\alpha\beta\left(1-\frac{\mu}{\gamma}-2m^*_\pm\right)
        -(\beta-\phi)\frac{\mu}{\gamma}-\mu\right],$$
so that
\begin{align*}
\det J_1({\textstyle\frac{\mu}{\gamma}},m^*_\pm)>0&\Leftrightarrow
  \left[\alpha\beta\left(1-\frac{\mu}{\gamma}-2m^*_\pm\right)
        -(\beta-\phi)\frac{\mu}{\gamma}-\mu\right]<0\\
&\Leftrightarrow m^*_\pm>\frac{1}{2}\left[\left(1-\frac{\mu}{\gamma}\right)
-\left(\frac{\beta-\phi}{\alpha\beta}\right)\frac{\mu}{\gamma}
-\frac{\mu}{\alpha\beta}\right]=-\frac{B}{2}.
\end{align*}
Since $m^*_-<-\frac{B}{2}<m^*_+$, we see that $E_4$ is never stable,
while the stability condition for $E_3$ reduces to
${\rm tr}\, J_1({\textstyle\frac{\mu}{\gamma}},m^*_+)<0$.  Thus we calculate
$${\rm tr}\, J_1({\textstyle\frac{\mu}{\gamma}},m^*_+)=
  (\beta-\phi)\left(1-2\frac{\mu}{\gamma}\right)-(\mu+\epsilon)
  -(\beta-\phi+\alpha\beta+\gamma)m^*_+,$$
so that
\begin{equation}
{\rm tr}\, J_1({\textstyle\frac{\mu}{\gamma}},m^*_+)<0\Leftrightarrow
m^*_+>\frac{(\beta-\phi)\left(1-2\frac{\mu}{\gamma}\right)-(\mu+\epsilon)}
           {(\beta-\phi+\alpha\beta+\gamma)}
     =\frac{(\beta-\phi)\,\frac{\mu}{\gamma}\,(R_2-2)}
           {(\beta-\phi)+\alpha\beta+\gamma}.
\label{E3LAS}
\end{equation}
This is true for all $m^*$ when $R_2<2$.  The case when $R_2\ge 2$ requires
further algebra and is relegated to the Appendix.  The result is that $E_3$
is always stable (LAS) when it exists, while $E_4$ is always unstable.

\subsection{Global Behavior}

%%% table 3 was here

Table \ref{equil1} summarizes the conditions for existence and local
stability of the four equilibria of the one-track model.  
Because system \eqref{S1}--\eqref{M1} can be reduced to a set of two
differential equations \eqref{v1b}--\eqref{m1b}, we can apply the
Poincar\'e-Bendixson Theorem to establish global stability.  Since the 
system is well-posed, with the invariant set $D=\{(v,m):\;v,m>0;\;v+m\le 1\}$,
there are no unbounded solutions beginning in the state space, and a
straightforward application of Dulac's Criterion with the function $b=1/vm$
confirms that there are no limit cycles, either:
$$\frac{\partial}{\partial v}\left(b\frac{dv}{dt}\right)
 +\frac{\partial}{\partial m}\left(b\frac{dm}{dt}\right)<0\;\;{\rm in}\;D.$$
Therefore, all solutions to the system which begin within the state
space must approach an equilibrium.  In particular, when only one locally
stable (LAS) equilibrium exists, that equilibrium must in fact be globally
stable (GAS).

\begin{figure}[!ht]
\begin{centering}
%\leavevmode
%\includegraphics[width = 0.7\textwidth, height = 0.7\textwidth]{r1r2.jpg}
\epsfxsize=0.7\textwidth \epsffile{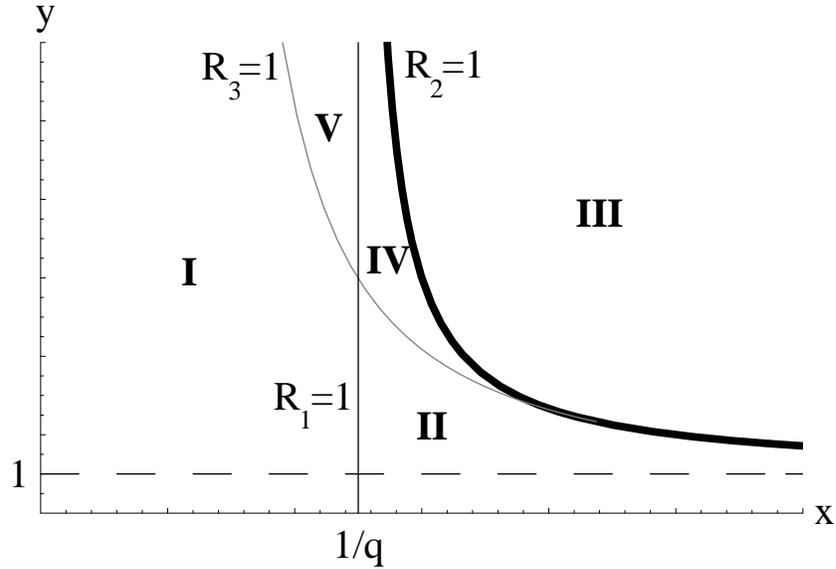} % was r1r2.eps
\caption{Regions of equilibrium stability, with $q$, $x$, $y$
as in Appendix~A} \label{r1r2}
\end{centering}
\end{figure}

% section explaining the five regions
The conditions in Table \ref{equil1} can be graphed to show the different
possible global behaviors of the one-track model.  Each condition corresponds
to a curve dividing parameter space into multiple regions, each of which
represents a different global behavior (see Appendix~\ref{blah} for the
derivations).  The five resulting regions are illustrated in Figure~\ref{r1r2}
and summarized in Table~\ref{stabreg}.

%%% table 4 was here

\begin{enumerate}
\item[I] In region I, $E_1$ is the only stable equilibrium; 
the party will always go extinct.
\item[II] In region II where $R_1>1$, $R_2<1$, and $R_3<1$,
%conditions \eqref{Bcond-main} and \eqref{discrimcond-main} are not met, 
$E_2$, the member-free state, is the only stable equilibrium.  
\item[III] In region III where both $R_1>1$ and $R_2>1$, $E_3$ is the only 
stable equilibrium; the party will inevitably approach a survival state.
\item[IV] In region IV where $R_1>1$, $R_2<1$ and $R_3>1$,
%conditions \eqref{Bcond-main} and \eqref{discrimcond-main} are met, 
$E_2$ and $E_3$ coexist as stable equilibria although, if placed
in a political context, $E_2$ is not a realistic outcome for the party.  
\item[V] In region V where $R_1<1$, $R_2<1$, and $R_3>1$,
%conditions \eqref{Bcond-main} and \eqref{discrimcond-main} are satisfied, 
$E_1$ and $E_3$ are both stable; depending on the initial conditions 
the solution tends to one state or the other.
\end{enumerate}

\begin{figure}[!ht]
\begin{centering}
%\leavevmode
%\includegraphics[width=0.7\textwidth, height=0.7\textwidth]{bifdiagram.jpg}
\epsfxsize=0.6\textwidth \epsffile{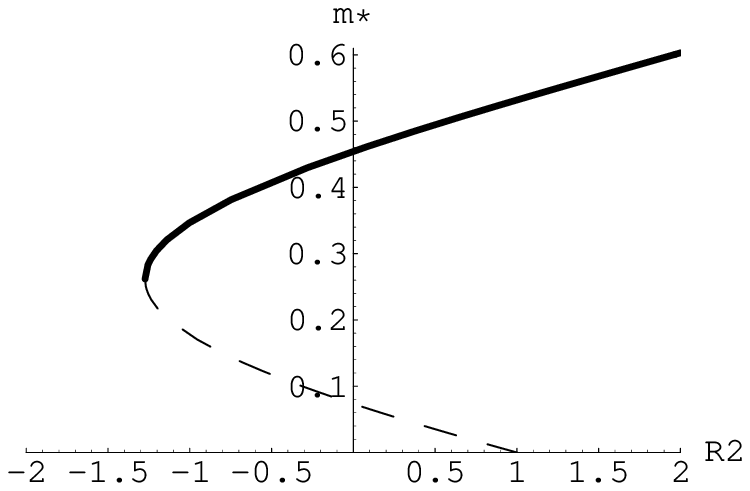}  % was bifdiagram.eps
\caption{Diagram illustrating a backward bifurcation that extends below $R_2=0$
 ($R_1=1$), using parameter values $\phi=0.15$, $\mu=0.05$, $\epsilon=0.25$,
 $\gamma=0.20$ (all rates in yr$^{-1}$), $\alpha=1.25$}
\label{bif}
\end{centering}
\end{figure}

% section explaining the backward bifurcation
The coexistence of two locally stable equilibria (one representing party
survival and the other party extinction) in regions IV and V involves a
phenomenon known in epidemic modeling as a backward bifurcation, in which
the exchange of stability at a threshold value (here, $R_2=1$) reverses
direction, creating a situation in which the unstable equilibrium $E_4$
serves to split the state space into two basins of attraction for the
two stable equilibria.  In these regions, the equilibrium approached
depends upon initial conditions: in particular, a large enough core of
dedicated members $M$ can sustain the party (toward $E_3$).
What is unusual here is that the ``backward'' part of the bifurcation curve 
can extend back beyond not only the bifurcation at $R_2=1$ but also the 
bifurcation at $R_1=1$ (where $R_2=0$), as illustrated in Figure~\ref{bif}.
Whenever $R_1<1$, $R_2<0$, a condition that would normally lead to 
the death of the party given local asymptotic stability of the PFE,
there are still conditions ($R_3>1$)
%(\eqref{Bcond-main} and \eqref{discrimcond-main})
under which two survival equilibria exist, one of which is stable ($E_3$).  
In other words, the party can thrive in conditions under which it 
would normally die out, given that we have the necessary parameters 
and sufficient initial number of $M$ individuals.  

The ideal conditions for the party, of course, are when $R_2>1$ (region~III),
in which the party survives regardless of initial conditions.
(Note $R_2>1$ implies $R_1>1$.)  The last part of this section interprets 
all these mathematical thresholds in political terms.

\subsection{Threshold Parameters $R_1$, $R_2$, and $R_3$}

Our system contains three local thresholds or tipping points where population
outcomes, measured as $S$, $V$, and $M$, depend on parameter values.
By tipping point we refer to the sociological term that describes
the point at which a stable phenomenon turns into a crisis, which, in
a political context, corresponds to the extreme states of the party:
death and growth 
%\cite{Gladwell}
(Gladwell, 2000).  In the context of our model, for
example, the party cannot sustain grassroots growth until parameter
conditions reach $R_1=1$, after which point the third party voting
and member classes gain individuals.  These threshold quantities
correspond to similar terms in demographic and epidemic models called
{\em reproductive numbers} which measure the average number of offspring
or infections caused by a single group member during its lifetime.
In our political context it is more appropriate to interpret these
reproductive numbers in collective terms as regards the influence of
party voters on the susceptible population.  We distinguish between the
aforementioned thresholds, $R_1$, $R_2$, and $R_3$, by analyzing them
qualitatively in a political context.

$R_1$ denotes the average number of susceptibles influenced to vote for
the party by a single party voter in $V$, if dropped into a homogeneous
population of susceptibles.  The expression
$R_1=\frac{\beta-\phi}{\mu+\epsilon}$ gives the net peer pressure, 
$\beta-\phi$, on susceptibles $S$ by voters $V$, multiplied by the 
average time, $\frac{1}{\mu+\epsilon}$, spent in the voting class $V$.  
As mentioned above, it is most appropriate to interpret $R_1$ as an
average number per voter taking into account the collective influence
of all the party voters.  Note that we are assuming that the influence
of party voters on susceptibles $\beta$ is stronger than the reverse influence
$\phi$ of susceptibles to discourage party voters, a necessary condition
for party emergence, which guarantees that $R_1>0$.

$R_2$ measures instead the second stage of recruitment: how effective 
party members $M$ are in recruiting third party voters to become members 
once there are enough individuals in $V$.  The expression
$R_2=\frac{\gamma}{\mu}(1-\frac{1}{R_1})$ gives the product of 
the rate at which party activists $M$ recruit voters from $V$ into $M$,
$\gamma$, the average political lifetime of a party member, $\frac{1}{\mu}$, 
and the proportion of the population $N$ in $V$ at the MFE,
$(1-\frac{1}{R_1})$ (since only voters $V$ can be recruited directly into $M$).
Similar to $R_1$, $R_2$ measures the average number of $V$ to $M$
conversions per individual in the $M$ class.  If party voters are ineffective
at influencing susceptibles to vote for the party and $R_1<1$, then the value
of $R_2$ will be negative; rather than interpreting this as a negative ability
of party members to recruit voters into activism, however, it should be seen
as an indication that the pool of party voters available for recruitment into
party membership is not there, preventing normal party growth.

$R_3$ measures the extent to which party members $M$ actively recruit
susceptible individuals $S$ into the voting class $V$.  This activism,
which sidesteps the traditional hierarchical structure of a party, is
a key characteristic of growing grassroots movements, which often lack
a political environment favorable to their growth in the traditional way
outlined above ($R_2<1$).  Because of the two conditions required
mathematically for party survival when $R_2<1$, $R_3$ is defined
as the greater of two quantities, both of which involve party members'
abilities to recruit voters, $r_2$, and members, $r_3$, over their 
political lifetimes, as well as the relative efficacy of voter 
peer-pressure influence to members' influence through activism 
on the susceptible, $q$.  The general form $r_3(1-1/r_2)$ present
in both components of $R_3$ parallels the form of $R_2=r_2(1-1/R_1)$,
with members' potential $r_2$ for converting third-party voters into members
reducing the constraint on their ability $r_3$ to produce those voters 
in the first place.  Note that $r_2$ must be great enough in order for
$R_3$ even to be positive: that is, only if the members can recruit
other members well can their recruitment of voters sustain the party.

We now list possible outcomes involving these thresholds and their implications:
\begin{enumerate}
\item When $R_1<1$ (regions I and V in Figure~\ref{r1r2} and
Table~\ref{stabreg}) the net influence of party voters $V$ upon susceptibles
is weak enough that grassroots emergence of the party is not possible.
The hierarchical structure of party involvement precludes normal growth
of party membership $M$ when party voters are unable to replenish their
own ranks ($R_1<1$ implies $R_2<0$).
However, exceptionally, if the recruiting ability of party members $M$
is great enough ($R_3>1$, region V), a sufficiently large dedicated core
can sustain the party through its own efforts, despite the relative
inefficiency or lack of influence of those merely voting for the party.
This outcome illustrates the key role activists play in party survival
during periods of adversity.

\item When $R_1>1$ each individual in $V$ is converting, on average, 
more than one person in $S$ into $V$, thus allowing the voting class
$V$ to thrive.  If, in addition, $R_2<1$ (regions II and IV), we have
a situation in which party activists are then unable to recruit
from the voting class effectively enough to maintain the party core $M$
(perhaps because the voting class is too small).  Normally this would
lead to an outcome in which a group disposed to vote for the party remains
(in $V$) but the party core itself dwindles away, leading effectively to
party extinction (with the ideologically closest main political party perhaps
adjusting a platform to capture these votes).  
However, as with the case when $R_1<1$, a sufficiently large initial core
can ensure the party's survival when the core is effective at influencing
individuals to begin voting for the party ($R_3>1$, region~IV).

\item The condition $R_2>1$ explains the case where the party ($V$
and $M$ classes) grows normally, by recruiting members from the $S$ and $V$
populations, respectively.  Here the party voters are influential enough
in garnering new voters from $S$ ($R_1>1$) that the party's survival does
not depend on party activists' ability to recruit new voters directly
(i.e., $R_3$ does not come into play).
%inequality \eqref{Bcond-main} need not be satisfied).
In epidemiological terms, this corresponds to a successful invasion:
conditions are so favorable for the development of the party that it
will become established even with a small initial group of members.
\end{enumerate}

% final discussion involving all 3 recruitment actions
In general, the survival of the party is determined by the interplay among
the three recruitment processes involved in the model: from $S$ to $V$ by
members of $V$, as measured by $R_1$; from $V$ to $M$ by members of $M$,
as measured by $r_2$ ($R_2$ incorporates both of these first two processes);
and from $S$ to $V$ by members of $M$, as measured by $r_3$ ($R_3$
%inequalities \eqref{Bcond-main} and \eqref{discrimcond-main} 
incorporates all three processes).  Party survival
requires either that the first two processes be effective enough
($R_1>1$ and $R_2>1$) for a small grassroots effort to take hold,
or else that an initial membership core be large enough, and the 
second and third processes effective enough ($R_3>1$),
%(inequalities \eqref{Bcond-main} and \eqref{discrimcond-main}), 
that direct recruitment by party activists can sustain its membership.

%%%%%%%%%%%%%%%%%%%%%%%%%%%%%%%%%%%%%%%%%%%%%%%%%%%%%%%%%%%%%%%%%%%%%%%%
%%             Section 4                                              %%
%%%%%%%%%%%%%%%%%%%%%%%%%%%%%%%%%%%%%%%%%%%%%%%%%%%%%%%%%%%%%%%%%%%%%%%%

\section{A Case Study: The Green Party}

\subsection{Methods}

As an application of the simplified one-track model analyzed in Section
\ref{sec3}, we used Green Party registration and voting records for
six states (CA, ME, MD, NY, OR, PA) and the District of Columbia (DC)
to study the growth of the Green Party during the past decade or so.

We began by establishing basic demographic information for the target
(study) population.  A report from the Pew Research Center (2007) %\cite{pew}
provided annual statistics on percentages of Americans identifying
themselves as confirmed Democrats or Republicans, leaning Democrat or
Republican, or confirmed or leaning toward third parties, for the period
1987--2007.  We took an average of these percentages during the period
2000--2007, during which time approximately 25\% of the voting population
identified itself as leaning Democrat or Republican, and an additional
about 11\% as confirmed or leaning toward third parties (including the
Green Party).  We estimate our target population---those capable of being
influenced to vote Green---as all of the former group, and about half of
the latter group (since the latter group also includes confirmed Greens).
Thus we estimate that the target population consists of about 25\%+6\%=31\%
of the total voting population of each state.  For each state, we averaged
the voting population size over the time period of interest (which varied
slightly from state to state, as detailed below) from voting records.
We then normalized the voting and registration records as proportions of
the target population in each case.

To determine the replacement (or mortality) rate $\mu$, we used the average
2003 life expectancy at birth in the U.S. of 77.5 years given in 
Shrestha (2006),
%\cite{crs},
and the minimum voting age of 18 years, to derive an average voting lifetime of
59.5 years, for an estimate of $\mu=1/59.5\,yr\approx 4.58\times 10^{-5}/day$.
(A United Nations (2008) report gives an average U.S. life expectancy at birth
of 78.3 years for the period 2000--2005, 
%\cite{unpop}
which yields a comparable estimate of $\mu=4.77\times 10^{-5}/day$.)

\begin{figure}
\centering
\epsfxsize=\textwidth\epsffile{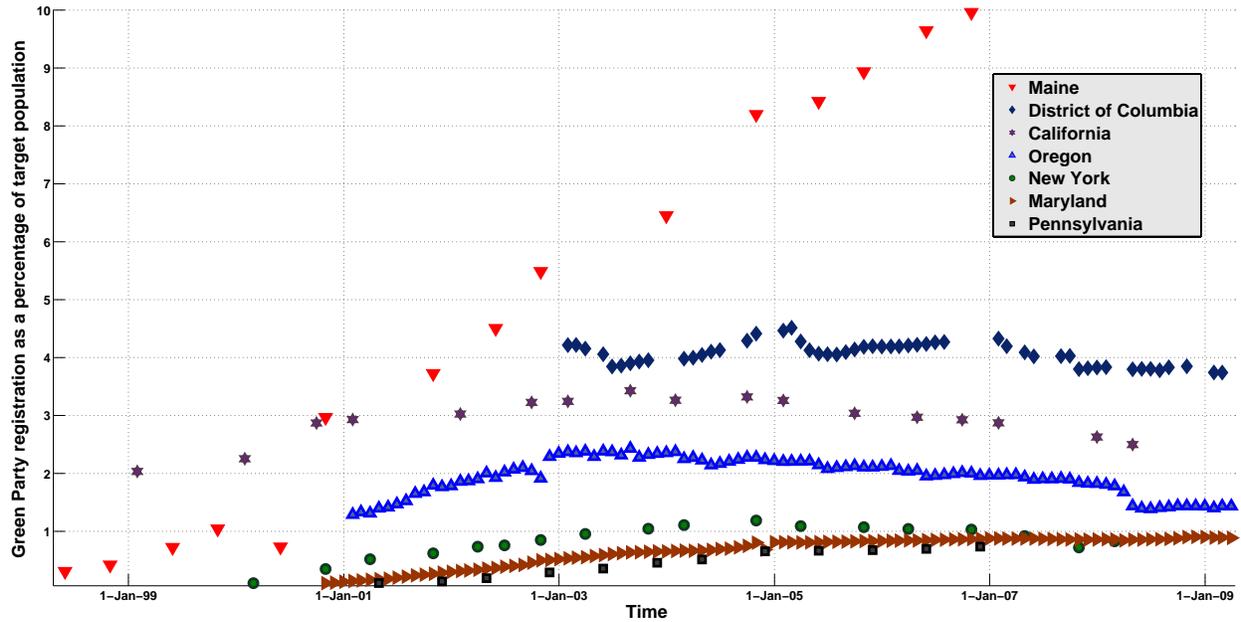}
\caption{Green Party registration data for the seven states (and district)
used in the case study, shown as percentages of the target population.}
\label{datafig}
\end{figure}

The data used for this case study came from official voter registration records
%\cite{CAdata,DCdata,MEdata,MDdata,NYdata,ORdata,PAdata} 
and election results
%\cite{CAvote,DCvote,MEvote,MDvote,NYvote,ORvote,PAvote} 
from each state (see bibliography for sources).
Voter registration records showing Green Party registration totals, given
in some states as often as monthly, were fit to the size of the member class
$M$ over time.  Since the dates for which Green Party registration data were
given varied from state to state, the initial and final times did as well,
but covered approximately the decade 1999--2008.  Since voting data were given
less often (in general in November of even-numbered years) these data were
used, where available, as initial conditions $V(0)+M(0)$.  In other cases
$V(0)$ was estimated along with other model parameters as discussed below.
Table~\ref{data1} gives the time periods modeled as well as the initial
conditions used.  The size of each class is given as a percentage of the
total target population (the size of which is also given in the table).
The rescaled Green Party registration data is also shown in
Figure~\ref{datafig}.
As can be seen in the graph, some states saw a noticeable change in the 
Green Party's trajectory following the November 2004 election, and so the
data for these states (DC, CA, OR, NY) was broken into two subseries (three
in the case of DC, which also underwent a visible change following the 2006
election), with different parameter estimates for each subseries.

%%% table 5 was here

Data fitting used the program Berkeley Madonna 8.3 to obtain a least-squares
fit to the Green Party registration data $M(t)$ for the model parameters
$\beta-\phi$, $\alpha\beta$, $\epsilon$, and $\gamma$, and, where necessary,
initial conditions.  ($\beta-\phi$ and $\alpha\beta$ were estimated directly,
rather than $\alpha$, $\beta$ and $\phi$ separately, in order to reduce the
number of free parameters.)  The estimation process was iterated until the
estimates were not on target interval boundaries, and this optimization was
carried out 10 times for each state (with different initial guesses) to obtain
the best possible fit.  The resulting estimates are also given in 
Table~\ref{data1}.  These estimates were then used to determine the state
of the system in each state by calculating reproduction numbers and predicted
end states (equilibria); results are shown in Table~\ref{data2}.

%%% table 6 was here

\subsection{Results}

As seen in Figure~\ref{datafig}, the data indicate a clear period of growth
for the Green Party in the selected states during the first five years of
the twenty-first century.  However, for some states (DC, CA, OR, NY) the data
also show a visible decrease beginning at the end of 2004.
The data indicate a clear change in the political landscape
in these states following the November 2004 election, in which the positive
trends the Green Party had been seeing in recent years reversed, and voters
left the Green Party, probably for the Democratic Party which saw gains in
the 2006 and 2008 elections.  The DC Greens rallied in 2005 but experienced
this same trend reversal following the 2006 election.  In each case, 
Table~\ref{data1} suggests that the primary reason for this change was 
a significant jump in the attention the target population paid to the media
and the activities of the two primary political parties, as evidenced by 
a marked (orders of magnitude) increase in the value of the parameter
$\epsilon$ (despite, in most cases, a simultaneous increase in the net peer
voter influence $\beta-\phi$).

The other three states in this study (ME, MD, PA), on the other hand,
have continued to see healthy sustained growth in the Green Party over
the past decade, although the growth in Maine appears nearly meteoric
compared to the slow building up in Maryland and Pennsylvania.
This difference is reflected in the much higher value of $R_1$ for Maine
(cf. Table~\ref{data2}).  The parameter estimates for Maine are much lower
than for Maryland (cf. Table~\ref{data1}), likely reflecting the lower
overall person-to-person contact rate (Maine has no large urban areas,
whereas Maryland's population is concentrated in them) as well as perhaps
lower sensitivity to others' opinions, but the ratios reflected in the
threshold quantities provide fertile ground for the Green Party's growth.
Our one-track model predicts (Table~\ref{data2}) a slightly higher equilibrium
level of proportional participation in MD than in ME, but it will take much
longer to reach that end state (during which time political influences may
change).

Pennsylvania, meanwhile, appears to be in a more unusual situation, as the
parameter estimates obtained from the data suggest the presence of a backward
bifurcation as discussed in Section~\ref{sec3}:  Here the tipping-point
thresholds ($R_1$, $R_2$) do not appear to favor the long-term survival
of the Green Party, but an initially large influx of Green voters would
allow the party to thrive, even enjoying majority status among the target
voting population.  The reason for this phenomenon (also seen in the initial
periods for DC) can be observed in the parameter estimates, which show a low
general sensitivity $\beta-\phi$ to the passive influence of community peers,
but a high sensitivity $\alpha\beta$ to grassroots activists (relative to
other PA parameters), so that a sufficiently high core of party activists
could sustain the party.  Despite its initial similarity to Maryland in terms
of the proportional data illustrated in Figure~\ref{datafig}, the growth of
the Pennsylvania Greens is indicated by the model to be part of a transient
response---it is just that, in this system, transient responses can last on the
order of decades, during which political influences can change significantly.

%%%%%%%%%%%%%%%%%%%%%%%%%%%%%%%%%%%%%%%%%%%%%%%%%%%%%%%%%%%%%%%%%%%%%%%%
%%             Section 5                                              %%
%%%%%%%%%%%%%%%%%%%%%%%%%%%%%%%%%%%%%%%%%%%%%%%%%%%%%%%%%%%%%%%%%%%%%%%%

\section{Analysis of the Two-Track Model}

Having thoroughly examined and interpreted the one-track model,
we now perform analysis on the original, heterogeneous two-track model.
Since the added complexity of this model precludes a complete qualitative
analysis, we shall make use of numerical analysis when necessary to
demonstrate behavior analogous to that exhibited by the simpler model.

First, we proportionalize the two-track model in the same way we did 
for the one-track model to get:

\begin{align}
\frac{dh}{dt}&=p\mu+\epsilon_H v_H+\phi_H(h+\sigma l)v_H
               -\beta_H(v_H+\sigma v_L+\alpha m)h-\mu h, \label{heqn} \\
\frac{dl}{dt}&=(1-p)\mu+\epsilon_L v_L+\phi_L(l+\sigma h)v_L
               -\beta_L(\sigma v_H+v_L+\alpha m)l-\mu l, \label{leqn} \\
\frac{dv_H}{dt}&=\beta_H(v_H+\sigma v_L+\alpha m)h-\epsilon_H v_H
               -\phi_H(h+\sigma l)v_H-\gamma m v_H-\mu v_H, \label{vheqn} \\
\frac{dv_L}{dt}&=\beta_L(\sigma v_H+v_L+\alpha m)l-\epsilon_L v_L
               -\phi_L(l+\sigma h)v_L-\gamma m v_L-\mu v_L, \label{vleqn} \\
\frac{dm}{dt}&=\gamma m v_H+\gamma m v_L-\mu m.\label{meqn}
\end{align}

%We begin by obtaining the party-free equilibrium (PFE) and the first 
%tipping point, $R'_1$, which is analogous to $R_1$ in the one track model.  
%We find that determining any other equilibria or thresholds analytically 
%is too complicated and we cannot extract anything politically relevant 
%from further analysis; therefore, we look at equilibria stability numerically
%by fixing parameter values in the Jacobian matrices.  We then present
%deterministic simulations to observe the outcomes of the various
%equilibrium conditions.  Finally, we offer bifurcation diagrams for
%the model and analyze the outcomes.

\subsection{The Party-Free Equilibrium (PFE) and $R'_1$}
%the Threshold/Tipping Point $R'_1$} 

The two-track model, like the simpler version, includes a party-free
equilibrium.  Observing that $v_H^*=v_L^*=m^*=0$ satisfies
$dv_H/dt=dv_L/dt=dm/dt=0$, we substitute into $dh/dt=dl/dt=0$ to
find the PFE; $E_1$ is $(p,1-p,0,0,0)$.  The stability of $E_1$ is
again tied to the first threshold quantity, which we shall denote
$R'_1$ (we shall use prime superscripts to denote thresholds for the
two-track model) and calculate using the next-generation operator method
%\cite{cccR0}
(e.g., Castillo-Ch\'avez, Feng and Huang, 2002), 
where $v_H$, $v_L$ and $m$ are considered the infective classes:
\begin{equation}
R'_1=\frac{1}{2}\left[r_{HH}+r_{LL}+\sqrt{(r_{HH}-r_{LL})^2
+4\,r_{LH}\,r_{HL}}\right],
\label{r1prime}
\end{equation}
where
$$r_{HH}=p\,\frac{\beta_H-\phi_H}{\mu+\epsilon_H+(1-p)\sigma\phi_H}
\;\;{\rm and}\;\;
  r_{LL}=(1-p)\,\frac{\beta_L-\phi_L}{\mu+\epsilon_L+p\sigma\phi_L},$$
respectively, are the high-affinity and low-affinity analogues of $R_1$
(that is, the influence of $v_H$ on $h$, and of $v_L$ on $l$), and
$$r_{LH}=p\,\frac{\beta_H\sigma}{\mu+\epsilon_L+p\sigma\phi_L}
\;\;{\rm and}\;\;
  r_{HL}=(1-p)\,\frac{\beta_L\sigma}{\mu+\epsilon_H+(1-p)\sigma\phi_H}$$
measure the cross-affinity influences (of $v_L$ on $h$ and $v_H$ on $l$).
See Appendix B for calculations.

$R'_1$ can be interpreted as the average number of individuals
converted into a third-party voter (in either voting class) 
by an individual in $v_H$ or $v_L$ introduced into a population where
no one yet votes for the given party.  The average is somewhat
complicated, as it involves four different contributing influences,
each represented by one of the four $r$'s in \eqref{r1prime}.
Each of these component numbers has the same form as $R_1$ for
the one-track model (q.v.), but measures of conversion of high-affinity
voters are multiplied by a factor of $p$, the proportion of the 
population which has a high affinity for the given party, while
measures of conversion of low-affinity voters are multiplied by
the proportion $1-p$ of low-affinity individuals.  In addition,
cross-affinity influences are reduced by the factor $\sigma$.
Note that in the extreme cases $p=0$ and $p=1$ $R'_1$ reduces to $R_1$.

We can further interpret the expression for $R'_1$ in political terms
by observing that
\begin{equation}
\max(r_{HH},r_{LL})<R'_1<\max(r_{HH},r_{LL})+\sqrt{r_{LH}\,r_{HL}}
\label{r1primebounds}
\end{equation}
(again see Appendix~B for details).  That is, $R'_1$ is at least as
great as each of the within-track conversion efficiencies, and exceeds
the maximum of the two (presumably $r_{HH}$) by less than the contribution
of cross-track influences.  This latter contribution is the geometric mean
of two terms representing a two-stage process, in which a voter in one track
converts an individual of the opposite affinity class into a third-party
voter, who then influences an individual in the first track to join the
voting class of the original voter, thereby completing the cycle.
Because this cross-affinity cycle has two stages, the appropriate measure
of its efficiency is a geometric mean of the two individual stages.
In the case that there is no cross-affinity influence ($\sigma=0$),
the tracks decouple completely at this stage, and \eqref{r1primebounds} 
reduces to $R'_1=\max(r_{HH},r_{LL})$.

The PFE is locally stable when $R'_1<1$, and unstable when $R'_1>1$.
In other words, each third-party voter introduced into a population
that includes high-affinity and low-affinity individuals must influence,
on average, more than one person to vote for the third party during
his/her voting lifetime, in order for the third-party voter classes
to persist, with the average defined by $R'_1$.

\subsection{The Member-Free Equilibrium (MFE) and $R'_2$}

The two-track model also has a second threshold parameter $R'_2$.  
Analogous to $R_2$ from the one-track model, we define $R'_2$ as 
the average number of third-party voters ($V_H$ and/or $V_L$) 
a member can convert into $M$ if introduced into a population of them.
Since $R'_2$ is primarily concerned with the transition from 
third-party voting to membership, we conveniently regard $M$ as 
the only infectious class, in order to apply the next-generation 
operator method to determine this threshold.  We then calculate,
from \eqref{meqn0} (or \eqref{meqn}):
$$\frac{\partial}{\partial M}\left(\frac{dM}{dt}\right)
 =\gamma(v_H+v_L)-\mu;$$
since this is scalar we seek simply the positive part divided by
the term that is subtracted (departing $M$):
$$R'_2=\frac{\gamma}{\mu}\left(v^\ast_{H2}+v^\ast_{L2}\right),$$
where $v^\ast_{H2}$ and $v^\ast_{L2}$ are the equilibrium values 
at the MFE.  This expression is analogous to that for the one-track model,
$$R_2=\frac{\gamma}{\mu}v^*=\frac{\gamma}{\mu}\left(1-\frac{1}{R_1}\right).$$

Solving explicitly for the MFE of the two-track model is complicated,
but we can show enough to suggest that, as expected, it is unique and
exists only for $R'_1>1$.  Since we have $m^*=0$, any MFE must be an
equilibrium of the subsystem
\begin{align}
\frac{dv_H}{dt}=&\beta_H(v_H+\sigma v_L)(p-v_H)
               -\phi_H(p-v_H+\sigma(1-p-v_L))v_H-(\mu+\epsilon_H)v_H, 
\label{vheqn2} \\
\frac{dv_L}{dt}=&\beta_L(\sigma v_H+v_L)(1-p-v_L)
               -\phi_L(1-p-v_L+\sigma(p-v_H))v_L-(\mu+\epsilon_L)v_L, 
\label{vleqn2}
\end{align}
since we can now rewrite $h=p-v_H$ and $l=1-p-v_L$.  The two resulting
equilibrium conditions can be simplified to a single equation of degree 4,
which admits up to 4 solutions.  One of these is the PFE, which can be
factored out to leave a cubic equation for the MFE.  It can be shown that
the constant term in this cubic equation is zero precisely when $R'_1=1$,
so that the number of positive solutions changes by one when $R'_1$ crosses 1.

In the special case that $\sigma=0$ (no cross-affinity influence),
the system \eqref{vheqn2}--\eqref{vleqn2} decouples, yielding equilibria
$E_1(0,0)$, $E_{2a}(\tilde{v}_H^*,0)$, $E_{2b}(0,\tilde{v}_L^*)$, and 
$E_{2c}(\tilde{v}_H^*,\tilde{v}_L^*)$, where 
$$\tilde{v}_H^*=p-\frac{\mu+\epsilon_H}{\beta_H-\phi_H}
  \;\;{\rm and}\;\;
  \tilde{v}_L^*=(1-p)-\frac{\mu+\epsilon_L}{\beta_L-\phi_L}$$
are meaningful only if positive,
i.e., if $r_{HH}>1$ and $r_{LL}>1$, respectively.  A straightforward
calculation of the Jacobian matrix shows that, within this subsystem,
\begin{itemize}
\item if $R'_1<1$ ($r_{HH}<1$ and $r_{LL}<1$), the only equilibrium is $E_1$,
  which is locally asymptotically stable (LAS);
\item if $r_{HH}>1$ and $r_{LL}<1$, $E_1$ is unstable but $E_{2a}$ is LAS;
\item if $r_{HH}<1$ and $r_{LL}>1$, $E_1$ is unstable but $E_{2b}$ is LAS;
\item if $r_{HH}>1$ and $r_{LL}>1$, $E_1$, $E_{2a}$ and $E_{2b}$ are all
  unstable but $E_{2c}$ is LAS.
\end{itemize}

Regardless of the value of $\sigma$, it is straightforward to show that
all solutions of \eqref{vheqn2}--\eqref{vleqn2} which begin within
$[0,p]\times[0,1-p]$ remain within those bounds, by observing that
$dv_H/dt<0$ when $v_H=p$ and $0\le v_L\le 1-p$, and that $dv_L/dt<0$ 
when $v_L=1-p$ and $0\le v_H\le p$.  One can also exclude limit cycles
from solutions of \eqref{vheqn2}--\eqref{vleqn2} under the assumptions
that $\beta_H>\phi_H$ and $\beta_L>\phi_L$, via the usual application
of Dulac's Criterion:
$$\frac{\partial}{\partial v_H}
  \left(\frac{1}{v_Hv_L}\,\frac{dv_H}{dt}\right)<0, \;\;
  \frac{\partial}{\partial v_L}
  \left(\frac{1}{v_Hv_L}\,\frac{dv_L}{dt}\right)<0.$$
We can therefore apply the Poincar\'e-Bendixson Theorem to conclude
that the equilibria of this system with $\sigma=0$ identified above
as locally stable, are in fact globally stable.

We can also differentiate the equilibrium conditions for
\eqref{vheqn2}--\eqref{vleqn2} implicitly by $\sigma$ to
see what happens as $\sigma$ increases from zero: for instance,
$$\frac{\partial v_L}{\partial\sigma}\Big|_{\sigma=0}
 =\frac{-\beta_Lv_H^*(1-p-v_L^*)+\phi_Lv_L^*(p-v_H^*)}
       {(\beta_L-\phi_L)(1-p-2v_L^*)-(\mu+\epsilon_L)},$$
so
$$\frac{\partial v_L}{\partial\sigma}\Big|_{E_{2a},\sigma=0}
 =\frac{-\beta_L\tilde{v}_H^*(1-p)}
       {(\beta_L-\phi_L)(1-p)-(\mu+\epsilon_L)}.$$
Thus when $r_{HH}>1$ and $r_{LL}>1$, the numerator is negative
and the denominator is positive, so that $E_{2a}$ exits the state space
as $\sigma$ increases from zero.  A similar calculation holds for $E_{2b}$.

Since \eqref{vheqn2}--\eqref{vleqn2} is a subsystem of
\eqref{heqn}--\eqref{meqn} (in which $m(t)\equiv 0$), stability in
the subsystem does not imply stability in the full system, but
instability in the subsystem does imply instability in the full system.
Thus the full system \eqref{heqn}--\eqref{meqn} has at most one stable
MFE when $\sigma=0$ (and, by continuity using the above result of implicit
differentiation, for $\sigma$ sufficiently small), that stability
depending upon the additional dimension ($m$) not present in
\eqref{vheqn2}--\eqref{vleqn2}, as measured by $R'_2$.
While $R'_2$ is not expressed explicitly, given the implicitness of
$v^\ast_{H2}$ and $v^\ast_{L2}$, the observed uniqueness of the MFE 
allows us to draw conclusions about the two-track model numerically.

\subsection{Survival Equilibria}

Although the equilibrium conditions for \eqref{heqn}--\eqref{meqn}
are too complicated to solve outright (apart from the fact that
$v_H^*+v_L^*=\mu/\gamma$ for any survival equilibrium, from \eqref{meqn}),
we can verify numerically not only their existence as expected when
$R'_2>1$, but also the existence, under certain conditions, of a
backward bifurcation at $R'_2=1$ just as observed for the one-track model.

\begin{figure}%[!b]

\begin{minipage}[t]{0.58\linewidth}
\centering
\epsfxsize=1.8in\epsffile{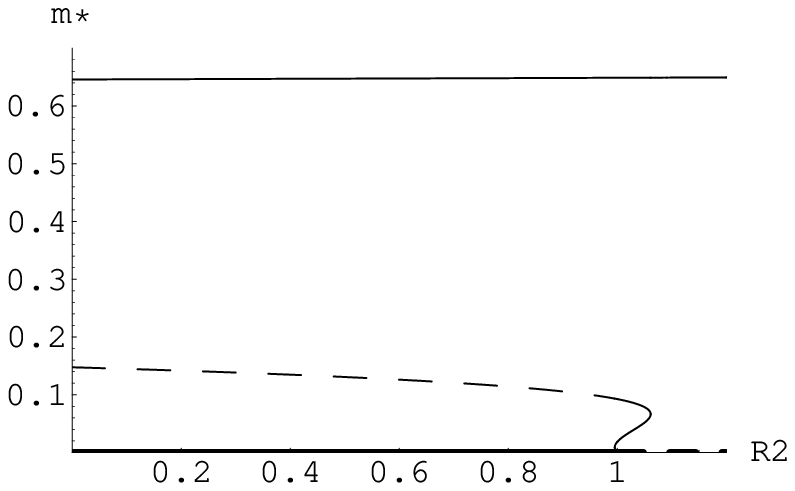}
\hfill
\epsfxsize=1.8in\epsffile{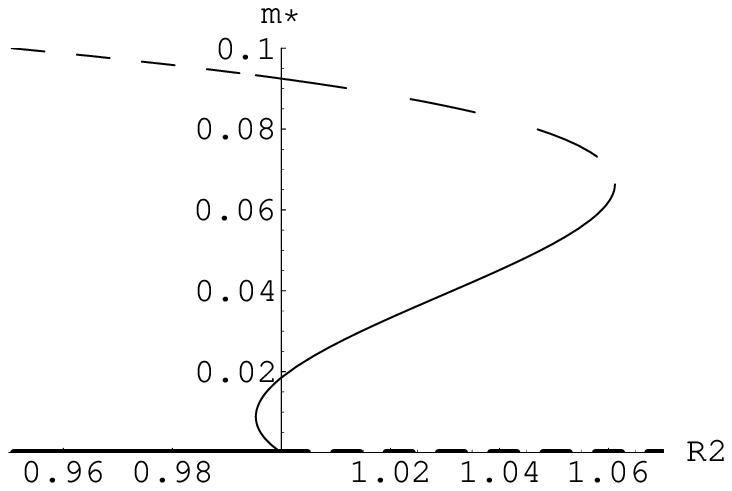}
\caption{A bifurcation diagram for the two-track model illustrating a backward
  bifurcation at $R'_2=1$ (see closer view at right), multiple saddle-node
  bifurcations, and survival equilibria extending below $R'_1=1$ ($R'_2=0$).
  Parameter values used are $p=0.1$, $\beta_L=0.2$, $\epsilon_H=0.25$,
  $\epsilon_L=0.5$, $\phi_H=\phi_L=0.1875$, $\alpha=1.25$,
  $\sigma=0.8$, $\gamma=0.6$, $\mu=0.05$; $\beta_H$ varies.  All rates
  are in years$^{-1}$ ($p$, $\alpha$, $\sigma$ are dimensionless).}
\label{2bbif1}
\end{minipage}
\hfill
\begin{minipage}[t]{0.38\linewidth}
\centering
\epsfxsize=2.2in\epsffile{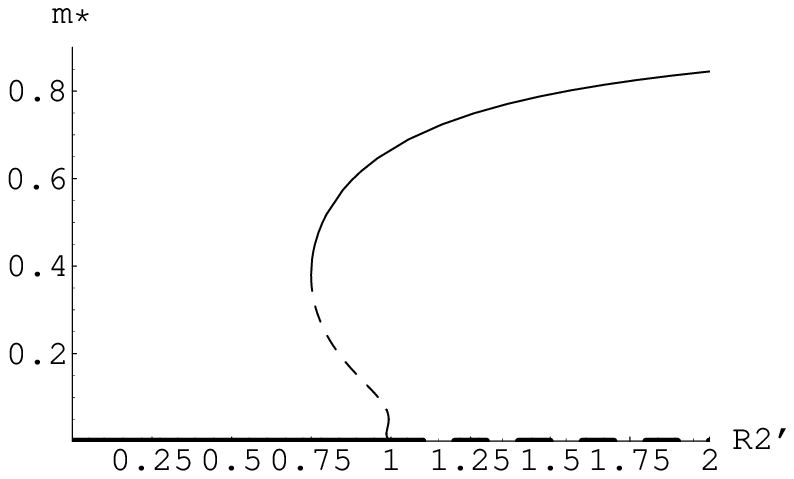}
\caption{A bifurcation diagram for the two-track model illustrating a backward
  bifurcation at $R'_2=1$.
  Parameter values used are $p=0.1$, $\beta_H=10$, $\beta_L=0.2$,
  $\epsilon_H=0.25$, $\epsilon_L=0.5$, $\phi_H=\phi_L=0.1875$, 
  $\alpha=1.25$, $\sigma=0.8$, $\mu=0.05$; $\gamma$ varies.  All rates
  are in years$^{-1}$ ($p$, $\alpha$, $\sigma$ are dimensionless).}
\label{2bbif2}
\end{minipage}

\end{figure}

Figure~\ref{2bbif1} shows a situation analogous to that depicted in
Figure~\ref{bif} for the one-track model, in which survival equilibria
may exist even below the PFE/MFE threshold ($R_2=0$ or $R'_2=0$);
in Figure~\ref{2bbif2} the critical (minimum) value of $R'_2$ is between
0 and 1.  Both figures, however, also demonstrate the existence of
multiple locally stable survival equilibria near $R'_2=1$ (see the 
close-up in Figure~\ref{2bbif1}), meaning that for some parameter values 
there are two different levels at which the party may stabilize, in addition 
to the stable (for $R'_2<1$) extinction equilibria PFE/MFE.  In these
situations, the initial number of party members plays a huge role in
determining whether the party surges to major growth, languishes, or
dies out entirely.

Although the parameter values used to create these figures are idealized
for illustration purposes, the value $p=0.1$ reflects an estimated 10\% of
the voting population having high affinity for the Green Party's agenda,
and other values reflect the distinctions in peer-driven behavior described
by the model for the two affinity classes.  In addition, similar curves
can be obtained using a wide range of values for model parameters.
Unfortunately, the complexity of the two-track model prevents an explicit
calculation of a quantity analogous to $R_3$ which measures the ability
of the member class to recruit ``susceptible'' voters from both affinity
classes to become third-party voters.  The same interpretation applies,
however: it is the work of party members in recruiting voters that enables
a party to persist when third-party voters' influence is too weak.

\section{Conclusion}

% remind the reader of the 2 main assumptions driving our models
The models described in this study investigate in mathematical terms
the consequences of our assumptions about the factors driving the growth
and persistence of a third political party arising through grassroots
efforts.  In particular, we assume a hierarchical structure in which
party members (activists) play a different, more extensive role in party
survival than party voters.  We also assume that the dominant influences
are primary contacts among third-party voters and members and the general 
public, manifested in our models as nonlinear terms involving the sizes
of the two groups making contact.  These nonlinearities govern the
behavior of the models---that is, the fate of the party under study---through
threshold quantities that describe the system's tipping points.
Our results should be taken as implications of the assumptions that
such primary contacts, and not other factors (apart from affinity
as defined for our two-track model), drive individuals' decisions
whether to support third political parties at any level.

% discuss/summarize threshold quantities
The primary result of our analysis is that our models identify, and
provide a way to measure, the three factors that determine the party's
survival.  Each of these factors is a reproductive number that describes
the ability of a given class in the party structure to recruit others.  
$R_1$ and $R'_1$ give the average number of unaffiliated voters recruited 
per third-party voter, for the simplified and general model, respectively.  
This first threshold quantity measures the voting class's ability to 
replace or sustain itself.  $R_2$ and $R'_2$, meanwhile, describes 
the efficiency with which third-party members convince third-party voters 
to become members.  This average number of new members recruited per 
existing member presupposes the success of the first stage of recruitment:
existing third-party voters recruiting new ones; that is, $R_1>1$, or else
$R_2<0$, which is meaningless except to indicate the failure of a
recruitment structure in which party members play no role in recruiting
new voters.  Finally, $R_3$ quantifies the ability of party members to
recruit new voters directly from the unaffiliated (with this party) public.
The growth of the Green Party in states like Maine and Maryland, and its
recent decline in states like California and Oregon, illustrate the effect
of these tipping-point thresholds (cf. Gladwell, 2000).
%\cite{Gladwell}

% interpret backward bifurcations
The model's prediction of survival states in scenarios where the primary,
hierarchical recruitment structure is not strong enough ($R_2<1$, and
even $R_1<1$) to allow a party to arise---that is, political trends are
not favorable and the resulting influence of peer pressure to vote for
and support a given party is weak---explains the persistence of established
third parties during periods of adverse conditions, when political winds
blow against them.  In particular, when party activists (members) are 
sufficiently capable of finding and recruiting new voters directly
(as measured by $R_3>1$), a large enough core of committed party
members can ensure the party's survival.  In cases where the political
environment was favorable for a time ($R_2>1$), this minimum core size is
typically reached quickly.  This reaching across traditional
hierarchical structures (rather than party members interacting primarily
with those who already vote for the party) provides a robustness to the
phenomenon that manifests mathematically in the backward bifurcations
illustrated in earlier sections.  Backward bifurcations also underscore
the importance of initial conditions (having enough initial party members)
in enabling that robustness.  This scenario is typified by the case study
of the Green Party in Pennsylvania.

% interpret effects of affinity classes
Our general model classifies the general voting population by affinity 
to the ideas and goals of a given third party.  The form of the expression
for $R'_1$ illustrates how a party's ability to take hold in even a small 
subset of the population (the high-affinity track) affects the party's
survival in the population as a whole: since $R'_1$ is greater than either
of the within-track voting replacement numbers (analogous to $R_1$),
a successful enough recruitment within the high-affinity track can maintain
the party.  Furthermore, the stratification into two tracks creates the
potential, as illustrated in Figures \ref{2bbif1} and \ref{2bbif2}, for
multiple stable survival states, which facilitate party growth since the
additional states require fewer initial members than the original one.
Our models could easily be extended to a stratification with intermediate
levels of affinity, each of which would make its own differential 
($R_1$-like) contribution to determining the overall recruitment potential
of a given third party (analogous to $R'_1$), in addition to increasing
the diversity of possible survival states.

%%% need to rewrite something here after addl. lit. review

% limitations and caveats
Even though we used available data to estimate parameters, we cannot claim
to have measured the strength of the various influences directly.  Our models
provide qualitative measures for the efficiency of parties' recruitment 
strategies, identifying which factors interact, and how.  The accuracy of 
these measures hinges, of course, on the underlying assumptions articulated
in Section~\ref{assume}.  Since it is often difficult in practice to
quantify the strength of another's opinions and arguments in influencing
one's opinion, our models can also be interpreted as an illustration
of how individual interactions within a population combine to exert
a single collective influence observable only at the population level.
Finally, it should be noted that the deterministic nature of our models
ignores the stochastic aspect of individual interactions (such as an
exceptionally charismatic individual): individual variability is a
critical factor when considering very small groups.  Here we have
described political behavior in terms of population averages, but
the first stages of any movement are entirely dependent on the particular
personalities involved.  Therefore our models should be considered
as picking up at the point where a grassroots movement has become
sufficiently organized to become a political force.

% inspirational quote
%Never doubt that a small group of thoughtful, committed citizens can change
%the world. Indeed, it's the only thing that ever has. ---Margaret Mead

%\newpage

\section*{Acknowledgements}
The authors gratefully acknowledge the participation of Karl Calderon,
Azra Panjwani, and Karen R\'ios-Soto in the research project
(and MTBI technical report) of which this paper is an extension.
The authors also thank Carlos Castillo-Ch\'avez, Linda Gao, 
Baojun Song, Armando Arciniega, Leon Arriola, and the participants
of MTBI 2005 for their help and advice.

This research has been partially supported by grants from the
National Security Agency, the National Science Foundation, the T
Division of Los Alamos National Lab (LANL), the Sloan Foundation,
and the Office of the Provost of Arizona State University.  The
authors are solely responsible for the views and opinions expressed
in this research; it does not necessarily reflect the ideas and/or
opinions of the funding agencies, Arizona State University, or LANL.

%%%%%%%%%%%%%%%%%%%%%%%%%%%%%%%%%%%%%%%%%%%%%%%%%%%
%%                         BIBLIOGRAPHY          %%
%%%%%%%%%%%%%%%%%%%%%%%%%%%%%%%%%%%%%%%%%%%%%%%%%%%

\section*{References}
\begin{hangref}

%\begin{thebibliography}{27}

%\bibitem{bet}
\item Bettencourt, L., Cintr\'on-Arias, A., Kaiser, D.I., \&
Castillo-Ch\'avez, C. (2006).  The power of a good idea: quantitative 
modeling of the spread of ideas from epidemiological models.
{\em Physica D}, 364, 513--536.
%LANL Technical Report LA-UR-05-0485, MIT-CTP-3589, 2005.

%\bibitem{brauer}
\item Brauer, F. \& Castillo-Ch\'avez, C. (2001).
{\it Mathematical models in population biology and epidemiology}.
New York: Springer-Verlag.

%\bibitem{CAvote}
\item California Secretary of State's Office (n.d.).
{\em Statewide election results}.  Online at \newline
http://www.sos.ca.gov/elections/elections\_elections.htm,
accessed 2009 July 1.

%\bibitem{CAdata}
\item California Secretary of State's Office (n.d.).
{\em Voter registration and participation statistics}.
Online at http://www.sos.ca.gov/elections/elections\_u.htm,
accessed 2009 July 1.

%\bibitem{cccR0}
\item Castillo-Ch\'avez, C., Feng, Z. \& Huang, W. (2002).  On the computation
  of $R_0$ and its role on global stability.  In C. Castillo-Chavez, S. Blower,
  P. Van den Driessche, D. Kirschner, \& A. Yakubu (Eds.), {\em Mathematical
  approaches for emerging and reemerging infectious diseases: an introduction}
  (IMA Vol. 125).  Berlin: Springer-Verlag.  pp. 224--250.

%\bibitem{CCC}
\item Castillo-Ch\'avez, C. \& Song, B. (2003).  Models for
the transmission dynamics of fanatic behaviors.  In H.T. Banks \&
C. Castillo-Ch\'avez (Eds.), {\it Bioterrorism: mathematical modeling
applications in homeland security} (SIAM Frontiers in Applied Mathematics 
Series).  Philadelphia: SIAM.  pp. 155-172. 

%\bibitem{crisosto}
\item Crisosto, N., Castillo-Ch\'avez, C., Kribs-Zaleta, C., \&
Wirkus, S. (2001).  Community resilience in collaborative learning.  
Cornell University Technical Report BU-1586-M.
%%% update publication citation

%\bibitem{DCvote}
\item District of Columbia Board of Elections and Ethics (n.d.).
{\em Election results}.
Online at http://www.dcboee.org/election\_info/election\_results/index.asp,
accessed 2009 July 1.

%\bibitem{DCdata}
\item District of Columbia Board of Elections and Ethics (n.d.).
{\em Voter registration statistics}.
Online at http://www.dcboee.org/voter\_stats/voter\_reg/voter.asp,
accessed 2009 July 1.

%\bibitem{Gladwell}
\item Gladwell, M. (2000).  {\it The Tipping Point.}
New York: Little, Brown and Company.

%\bibitem{Gonzalez}
\item Gonz\'alez, B., Huerta-S\'anchez, E., Kribs-Zaleta, C., 
Ortiz-Nieves, A., \& V\'azquez-Alvarez, T. (2003).  Am I too fat?  Bulimia
as an epidemic.  {\it Journal of Mathematical Psychology}, 47(5--6), 515--526.

%\bibitem{grano}
\item Granovetter, M. (1978).  Threshold models of collective behavior.
{\em American Journal of Sociology}, 83(6), 1420--1443.

%\bibitem{dmr1}
\item Huckfeldt, R. \& Sprague, J. (1995).  {\em Citizens, politics,
  and social communication: information and influence in an election
  campaign}.  New York: Cambridge University Press.

%\bibitem{kowal}
\item Kowalewski, D. (1995).  How movements move: the dynamics of an
ecoprotest campaign.  {\it The Social Science Journal}, 32(1), 49--67.

%\bibitem{macy}
\item Macy, M.W. (1991).  Threshold effects in collective action.
  {\em American Sociological Review}, 56(6), 730--747.

%\bibitem{MEvote}
\item Maine Bureau of Corporations, Elections \& Commissions (n.d.).
{\em Election results}.
Online at http://maine.gov/sos/cec/elec/prior1st.htm, accessed 2009 July 1.

%\bibitem{MEdata}
\item Maine Bureau of Corporations, Elections \& Commissions (n.d.).
{\em Voter registration}.
Online at http://www.maine.gov/sos/cec/elec/votreg.htm, accessed 2009 July 1.

%\bibitem{marks}
\item Marks, J. (1997).  A historical look at Green structure: 
1984 to 1992.  {\em Synthesis/Regeneration}, 14.  Online
at http://www.greens.org/s-r/14/14-03.html, accessed 2007 July 1.

%\bibitem{MDvote}
\item Maryland State Board of Elections (n.d.).
{\em Elections by year}.
Online at \newline http://elections.state.md.us/elections/index.html,
accessed 2009 July 1.

%\bibitem{MDdata}
\item Maryland State Board of Elections (n.d.).
{\em Monthly voter registration activity reports}.
Online at http://elections.state.md.us/voter\_registration/monthly.html,
accessed 2009 July 1.

%\bibitem{NYdata}
\item New York State Board of Elections (n.d.).
{\em Enrollment by county}.
Online at \newline http://www.elections.state.ny.us/EnrollmentCounty.html,
accessed 2009 July 1.

%\bibitem{NYvote}
\item New York State Board of Elections (n.d.).
{\em Election results}.
Online at \newline http://www.elections.state.ny.us/ElectionResults.html,
accessed 2009 July 1.

%\bibitem{dmr2}
\item Nickerson, D.W. (2008).  Is voting contagious?  Evidence
  from two field experiments.  {\em American Political Science Review},
  102, 49--57.

%\bibitem{olmar}
\item Oliver, P.E. \& Marwell, G. (1988).  The paradox of group
  size in collective action: a theory of critical mass, II.  {\em American
  Sociological Review}, 53, 1--8.

%\bibitem{ORdata}
\item Oregon Secretary of State Election Division (n.d.).
{\em Election registration and participation history}.
Online at http://www.sos.state.or.us/elections/votreg/regpart.htm,
accessed 2009 July 1.

%\bibitem{ORvote}
\item Oregon Secretary of State Election Division (n.d.).
{\em Elections history}.
Online at \newline http://www.sos.state.or.us/elections/other.info/stelec.htm,
accessed 2009 July 1.

%\bibitem{PAdata}
\item Pennsylvania Bureau of Commissions, Elections \& Legislation (n.d.).
{\em Voter registration statistics archives}.
Online at http://www.dos.state.pa.us/elections/cwp/view.asp? \newline
a=1310\&q=447072,
accessed 2009 July 1.

%\bibitem{PAvote}
\item Pennsylvania Bureau of Commissions, Elections \& Legislation (n.d.).
{\em Elections information}.
Online at
http://www.electionreturns.state.pa.us/ElectionsInformation.aspx?Function
\newline ID=0,
accessed 2009 July 1.

%\bibitem{pew}
\item The Pew Research Center for the People \& the Press (2007).
{\em Trends in political values and core attitudes: 1987-2007.
Political landscape more favorable to Democrats}.  Washington, DC: Author.
Online at http://people-press.org/reports/pdf/312.pdf, accessed 2009 July 1.

%\bibitem{unpop}
\item Population Division of the Department of Economic and Social
Affairs of the United Nations Secretariat (2008).  {\em World population
prospects: the 2008 revision}.  Online at http://esa.un.org/unpp, accessed
2009 July 1.

%\bibitem{dmr3}
\item Robinson, J.P. (1976).  Interpersonal influence in election campaigns:
  two step-flow hypotheses.  {\em Public Opinion Quarterly}, 40, 304--319.

%\bibitem{crs}
\item Shrestha, L.B. (2006).  {\em Life expectancy in the United States}
(updated August 16, 2006).  Washington, DC: Congressional Research Service,
Library of Congress.  Online at \newline http://aging.senate.gov/crs/aging1.pdf,
accessed 2009 July 1.

%\bibitem{X}
\item Song, B., Castillo-Garsow, M., R\'ios-Soto, K.R., Mejran, M.,
Henson, L., \& Castillo-Ch\'avez, C. (2006).  Raves, clubs and ecstasy: 
the impact of peer pressure.  {\it Mathematical Biosciences and Engineering},
3(1), 249--266.

%\bibitem{southwell1}
\item Southwell, P.L. (2003).  The politics of alienation: 
nonvoting and support of third-party candidates among 18--30-year-olds.
{\it The Social Science Journal}, 40(1), 99--107.

%\bibitem{southwell2}
\item Southwell, P.L. (2004).  Nader voters in the
2000 Presidential Election: what would they have done without him?
{\it The Social Science Journal}, 41(3), 423--431.

%\bibitem{timpone}
\item Timpone, R.J. (1998).  Ties that bind: measurement, demographics, 
and social connectedness.  {\it Political Behavior}, 20(1), 53--77.

%\bibitem{supreme}
\item Warren, E. (1957).  U.S. Supreme Court plurality opinion,
Sweezy v. New Hampshire, 354 U.S. 234, 250--251.

%\bibitem{wong}
\item Wong, J. (2000).  The effects of age and political exposure on the 
development of party identification among Asian American and Latino 
immigrants in the United States.  {\it Political Behavior}, 22(4), 341--371.

\end{hangref}

\newpage

\appendix
\section{Equilibrium Analysis for the One-Track Model}
\subsection{Proof of Proposition 1: Existence of $E_3$ and $E_4$}\label{blah}

We begin with the survival equilibrium condition $f(m^*)=m^{*2}+BM^*+C=0$,
where (from \eqref{BandC})
$$B=\left(\frac{\beta-\phi}{\alpha\beta}\right)\frac{\mu}{\gamma}     
-\left(1-\frac{\mu}{\gamma}\right)+\frac{\mu}{\alpha\beta}, \;\; 
C=-\frac{\mu}{\gamma}\left[\frac{\beta-\phi}{\alpha\beta}
\left(1-\frac{\mu}{\gamma}\right)-\frac{\mu+\epsilon}{\alpha\beta}\right].$$
We rewrite these expressions by defining the following terms:
$$q=\frac{\beta-\phi}{\alpha\beta}, \;\; r=\frac{\mu}{\mu+\epsilon}, \;\;
  x=\frac{\alpha\beta}{\mu+\epsilon}, \;\; y=\frac{\gamma}{\mu}.$$
By assumption, $\beta>\phi$, $\gamma>\mu$, and $\alpha>1$, so that
$0<q<1$, $0<r<1$, and $y>1$.  In these terms, $R_1=qx$, 
$R_2=y\left(1-\frac{1}{qx}\right)$, and
$$B=\frac{1+q}{y}-\left(1-\frac{r}{x}\right), \;\;
  C=-\frac{q}{y}\left(1-\frac{1}{y}-\frac{1}{qx}\right).$$

In order to be meaningful, solutions must fall within the interval
$(0,1-v^*)$ (since $m^*>1-v^*$ will make $s^*=1-v^*-m^*<0$).  We calculate
$$f(1-v^*)=f\left(1-\frac{1}{y}\right)=\frac{r}{x}+\frac{1-r}{xy}>0$$ and
$$f'(1-v^*)=2(1-v^*)+B=1-\frac{1-q}{y}+\frac{r}{x}>0 \;\;
  ({\rm since}\;\;1-q<1<y).$$
These two inequalities imply that any roots of $f$ lie to the left of $1-v^*$.
Next note that $f(0)=C=-\frac{q}{y^2}(R_2-1)$.  Then when $R_2>1$, $f(0)<0$, 
so there is exactly one solution $m^*_+$ in $(0,1-v^*)$.  When $R_2<1$, 
$f(0)>0$, so the number of solutions in $(0,1-v^*)$ is even.  In this case,
both solutions $m^*_\pm$ are in $(0,1-v^*)$ precisely when $B^2-4C\ge 0$ (the
solutions are real) and $B<0$ (the parabola's vertex $m=-B/2$ lies to the
right of 0).

The condition $B<0$ can be shown equivalent to $x>r$ and
$y>\hat{y}_B\equiv (1+q)\frac{x}{x-r}$.  The condition $C>0$ (i.e., $R_2<1$)
can be shown equivalent to $qx<1$ or $y<\hat{y}_C\equiv\frac{qx}{qx-1}$.
The condition $B^2-4C\ge 0$ becomes
$$\left(\frac{1+q}{y}-\frac{x-r}{x}\right)^2
  +4\frac{q}{y}\left(1-\frac{1}{y}-\frac{1}{qx}\right)\ge 0,$$
and multiplying by $x^2y^2$ we get
$$(x-r)^2y^2-2x\left[(1-q)(x-r)+2(1-qr)\right]y+(1-q)^2x^2\ge 0.$$
Solving the quadratic inequality in $y$, this means $y$ must not be between
the two positive roots
$$\hat{y}_\pm\equiv \frac{x}{(x-r)^2}\left\{\left[(1-q)(x-r)+2(1-qr)\right]
\pm 2\sqrt{(1-qr)\left[(1-q)(x-r)+(1-qr)\right]}\right\}.$$
Thus in order to have two survival equilibria, we must have $x>r$,
$\hat{y}_B<y<\hat{y}_C$, and either $y\le\hat{y}_-$ or $y\ge\hat{y}_+$.

In order to simplify these criteria, we compare the threshold values for $y$.
We find that $\hat{y}_-<\hat{y}_B<\hat{y}_+$ is equivalent (after
substitution) to
\begin{equation}
-\sqrt{(1-qr)\left[(1-q)(x-r)+(1-qr)\right]}<qx-1
<\sqrt{(1-qr)\left[(1-q)(x-r)+(1-qr)\right]},\label{critorder}
\end{equation}
or simply $|qx-1|<\sqrt{(1-qr)\left[(1-q)(x-r)+(1-qr)\right]}$.
Squaring both sides and simplifying leads to the compound inequality
$$r<x<\hat{x}_+\equiv\frac{1+q(1-r)}{q^2}.$$
The lower bound $r$ corresponds to the vertical asymptote in $x$ shared by
$\hat{y}_\pm$ and $\hat{y}_B$.  Since $q<1$,
$\frac{1}{q}<\frac{1}{q^2}<\hat{x}_+$, so at the upper bound $qx-1>0$,
and it is the second inequality in \eqref{critorder} that is violated.
That is, when $x=\hat{x}_+$, $\hat{y}_B=\hat{y}_+$.  Thus it is always true
that $\hat{y}_-<\hat{y}_B$, so the criterion $y>\hat{y}_B$ allows us to
discard the condition $y\le\hat{y}_-$ in favor of $y\ge\hat{y}_+$.

\begin{figure}
\centering
\epsfysize=3in\epsffile{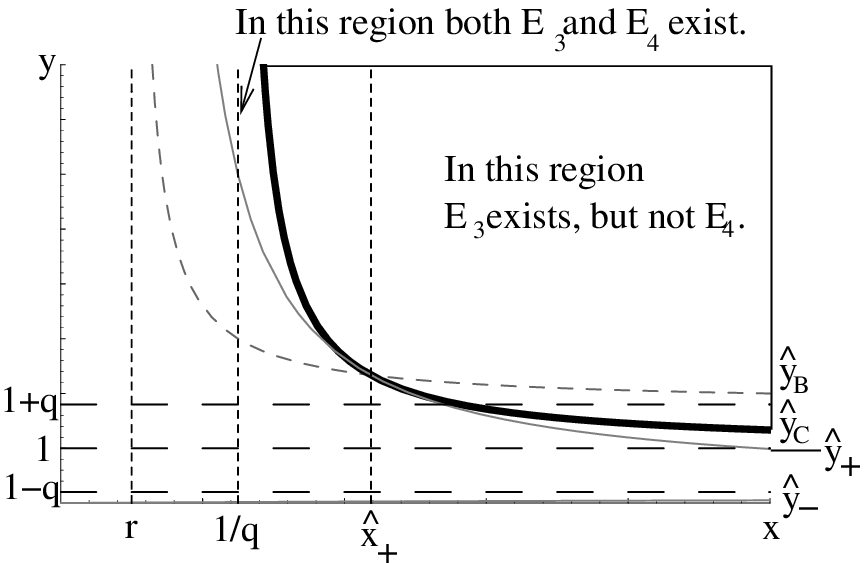}  % was 4curves.eps
\caption{Four threshold curves illustrating the survival equilibrium
  conditions}
\label{4curves}
\end{figure}

If we solve the inequality $\hat{y}_B<\hat{y}_C$ for $x$ (when $qx>1$), 
we find again the condition $x<\hat{x}_+$, indicating that the graphs of
$\hat{y}_B$, $\hat{y}_C$, and $\hat{y}_+$ all cross at $x=\hat{x}_+$.
Further similar computation can show that $\hat{y}_+<\hat{y}_C$ except
at $x=\hat{x}_+$, where they are tangent.  We thus require $r<x<\hat{x}_+$
and $\hat{y}_+\le y<\hat{y}_C$ (the latter inequality only for $qx>1$).
A graph illustrating all four curves is given in Figure~\ref{4curves}.

To put the survival equilibrium conditions back in terms of the original
parameters, we see that $r<x<\hat{x}_+$ becomes
$$\frac{\mu}{\mu+\epsilon}<\frac{\alpha\beta}{\mu+\epsilon}<
  \frac{1+\frac{\beta-\phi}{\mu+\epsilon}\,\frac{\epsilon}{\mu+\epsilon}}
       {\left(\frac{\beta-\phi}{\mu+\epsilon}\right)^2};$$
each of the inequalities can be solved for $\alpha\beta/\mu$ to make
$$\frac{\alpha\beta}{\mu}>\max\left(1,
  \frac{\beta-\phi}{\mu+\epsilon}\,\frac{\beta-\phi-\epsilon}{\mu}\right).$$
Note that the last expression in the inequality above is the product of $R_1$
and another fraction whose value exceeds 1 precisely when $R_1$ does.
Thus if $R_1<1$, the condition is simply $\frac{\alpha\beta}{\mu}>1$,
while if $R_1>1$ the condition is $\frac{\alpha\beta}{\mu}>
R_1(\beta-\phi-\epsilon)/\mu$.
It is not so simple to rewrite $y>\hat{y}_+$; however, we can return to the
original conditions $B<0$, $B^2-4C\ge 0$, which can be rewritten more simply,
either as $R_{3a}>1$, $R_{3b}>1$ as in the statement of Proposition~1, or as:
\begin{align}
B<0\Leftrightarrow & \frac{1}{\alpha\beta}
+\left(1+\frac{\beta-\phi}{\alpha\beta}\right)\frac{1}{\gamma}<\frac{1}{\mu},
\label{Bcond}\\
B^2-4C\ge 0\Leftrightarrow & \sqrt{\frac{1}{\alpha\beta}}
+\sqrt{\left(1-\frac{\beta-\phi}{\alpha\beta}\right)\frac{1}{\gamma}
+h(\epsilon/\gamma)}\le\sqrt{\frac{1}{\mu}},
\label{discrimcond}
\end{align}
where
$$h(\epsilon/\gamma)=2\sqrt{\frac{1}{\alpha\beta}}\sqrt{\frac{1}{\mu}}
\left(\sqrt{1+\frac{\epsilon}{\gamma}}-1\right)>0.$$

\bigskip

In order to rewrite the condition $B^2-4C\geq0$ in the form
\eqref{discrimcond}, we substitute $B$ and $C$ from \eqref{BandC} and get:
$$\left[\frac{\mu}{\alpha\beta}+\left(1+\frac{\beta-\phi}{\alpha\beta}\right)
\frac{\mu}{\gamma}-1\right]^2-4\frac{\mu}{\gamma}\left[
\frac{\mu+\epsilon}{\alpha\beta}-\frac{\beta-\phi}{\alpha\beta}
\left(1-\frac{\mu}{\gamma}\right)\right]\geq0.$$
Expansion, summing like terms, and completing the square yields
$$\left[\frac{\mu}{\alpha\beta}-\left(1-\frac{\beta-\phi}{\alpha\beta}\right)
\frac{\mu}{\gamma}+1\right]^2\geq
4\frac{\mu}{\alpha\beta}\left(1+\frac{\epsilon}{\gamma}\right).$$
Next we take the square root of both sides,
$$\left|\frac{\mu}{\alpha\beta}-\left(1-\frac{\beta-\phi}{\alpha\beta}\right)
\frac{\mu}{\gamma}+1\right|\geq
2\sqrt{\frac{\mu}{\alpha\beta}\left(1+\frac{\epsilon}{\gamma}\right)}.$$
Since we require $\mu<\gamma$ and $\beta>\phi$, then
$$\frac{\mu}{\alpha\beta}-\left(1-\frac{\beta-\phi}{\alpha\beta}\right)
\frac{\mu}{\gamma}+1=\frac{\mu}{\alpha\beta}+\frac{\beta-\phi}{\alpha\beta}
\frac{\mu}{\gamma}+\left(1-\frac{\mu}{\gamma}\right)>0,$$
so we can drop the absolute value bars.  Finally, rearranging, we get
$$\left(1-\frac{\beta-\phi}{\alpha\beta}\right)\frac{1}{\gamma}
\le\frac{1}{\mu}-2\sqrt{\frac{1}{\alpha\beta}\,\frac{1}{\mu}\,
\left(1+\frac{\epsilon}{\gamma}\right)}+\frac{1}{\alpha\beta}.$$
In the case that $\epsilon=0$, this inequality can be further
simplified by factoring the right-hand side as a perfect square, taking the
square root of both sides, and using the fact that \eqref{Bcond} implies
$1/\alpha\beta<1/\mu$, to get
$$\sqrt{\frac{1}{\alpha\beta}}+\sqrt{\left(1-\frac{\beta-\phi}{\alpha\beta}
\right)\frac{1}{\gamma}}\le\sqrt{\frac{1}{\mu}}.$$
We can apply the same technique to obtain \eqref{discrimcond}.

\subsection{Stability Analysis for $E_3$}

From the end of Section \ref{E3E4sec}, it remains to show that 
${\rm tr}\, J_1(E_3)<0$ when $R_2\ge 2$.  This condition on the trace
is equivalent to
$$m^*_+>L\equiv
  \frac{(\beta-\phi)\left(1-2\frac{\mu}{\gamma}\right)-(\mu+\epsilon)}
       {(\beta-\phi+\alpha\beta+\gamma)}.$$
In terms of $q$, $r$, $x$ and $y$ as defined in the previous section,
$$L=\frac{q\left(1-\frac{1}{qx}-\frac{2}{y}\right)}
         {\left(1+q+\frac{yr}{x}\right)}, \;\;
  B=\frac{\left(1+q+\frac{yr}{x}\right)}{y}-1, \;\;
  C=-\frac{q}{y}\left(1-\frac{1}{qx}-\frac{1}{y}\right).$$
Now the stability condition is
\begin{align*}
m^*_+=\frac{1}{2}\left[-B+\sqrt{B^2-4C}\right]&>L\\
\sqrt{B^2-4C}&>B+2L\\
B+2L<0,\;\;{\rm or}\;\;B+2L>0\;{\rm and}\;B^2-4C&>B^2+4BL+4L^2\\
B+2L>0\;\Rightarrow\;-C&>BL+L^2\\
B+2L>0\;\Rightarrow\;\frac{q}{y}\left(1-\frac{1}{qx}-\frac{1}{y}\right)&
>\frac{q}{y}\left(1-\frac{1}{qx}-\frac{2}{y}\right)-L+L^2\\
B+2L>0\;\Rightarrow\;0&>-\frac{q}{y^2}-L+L^2
\end{align*}
Here we multiply by $-\left(1+q+\frac{yr}{x}\right)^2/q$, expand, and
simplify to get
$$0<\left(\frac{1+q}{y}+\frac{r}{x}\right)^2
   +\left(1-\frac{1}{qx}-\frac{2}{y}\right)
    \left(1+\frac{1+yr}{x}+\frac{2q}{y}\right),$$
which is true since
$$R_2\ge 2\Leftrightarrow \left(1-\frac{1}{qx}-\frac{2}{y}\right)\ge 0.$$
This completes the verification that $E_3$ is LAS when it exists.

\section{$R'_1$ for the Two-Track Model}

We calculate the reproductive number $R'_{1}$ of the two-track model
using the next-generation operator method where $R'_{1}$ is
analogous to $R_1$ of the one-track model.  The party-free
equilibrium $E_1$ of \eqref{heqn}--\eqref{meqn} is $(p,(1-p),0,0,0)$.  
Differentiating \eqref{vheqn}--\eqref{meqn} with respect to the 
``infective'' variables $v_H$, $v_L$, and $m$, and substituting the PFE
values yields the following ``mini-Jacobian'' matrix:
%% unsubstituted version
%$$
%A=\begin{pmatrix}
%(\beta_H-\phi_H)h^*-(\mu+\epsilon_H+\phi_H\sigma l^*)& 
%\beta_H\sigma h^* &
%\alpha\beta_H h^*-\gamma v_H^*\\
%\beta_L\sigma l^* &
%(\beta_L-\phi_L)l^*-(\mu+\epsilon_L+\phi_L\sigma h^*)&
%\alpha\beta_L l^*-\gamma v_L^*\\
%\gamma m^* & \gamma m^* & \gamma(v_H^*+v_L^*)-\mu\\
%\end{pmatrix}
%$$
$$
A=\begin{pmatrix}
(\beta_H-\phi_H)p-w_H& 
\beta_H\sigma p &
\alpha\beta_H p \\
\beta_L\sigma(1-p) &
(\beta_L-\phi_L)(1-p)-w_L&
\alpha\beta_L(1-p) \\
0 & 0 & -\mu\\
\end{pmatrix},
$$
where $w_H=\mu+\epsilon_H+\phi_H\sigma(1-p)$ 
and $w_L=\mu+\epsilon_L+\phi_L\sigma p$.
We next rewrite $A=\tilde{M}-\tilde{D}$, where the entries of $\tilde{M}$ 
are nonnegative and $\tilde{D}$ is a diagonal matrix:
$$
\tilde{M}=\begin{pmatrix}
(\beta_H-\phi_H)p & \beta_H\sigma p & \alpha\beta_H p \\
\beta_L\sigma(1-p) & (\beta_L-\phi_L)(1-p)& \alpha\beta_L(1-p) \\
0 & 0 & 0 \\
\end{pmatrix}
$$
and
$$
D=\begin{pmatrix}
\mu+\epsilon_H+\phi_H\sigma(1-p)&0&0\\
0&\mu+\epsilon_L+\phi_L\sigma p&0\\
0& 0& \mu\\
\end{pmatrix}
$$
Now $R'_1$ is the dominant (largest) eigenvalue of
$$
MD^{-1}=\begin{pmatrix}
\frac{(\beta_H-\phi_H)p}{\mu+\epsilon_H+\phi_H\sigma(1-p)}&
\frac{\beta_H\sigma p}{\mu+\epsilon_L+\phi_L\sigma p}& 
\alpha\beta_H p/\mu \\
\frac{\beta_L\sigma(1-p)}{\mu+\epsilon_H+\phi_H\sigma(1-p)}&
\frac{(\beta_L-\phi_L)(1-p)}{\mu+\epsilon_L+\phi_L\sigma p}& 
\alpha\beta_L(1-p)/\mu \\
0&0&0\\
\end{pmatrix}
=\begin{pmatrix}
r_{HH} & r_{LH} & \alpha\beta_H p/\mu \\
r_{HL} & r_{LL} & \alpha\beta_H p/\mu \\
0&0&0\\
\end{pmatrix}
.
$$
The three eigenvalues are 0 and
$$\frac{1}{2}\left[r_{HH}+r_{LL}\pm\sqrt{(r_{HH}-r_{LL})^2+4r_{HL}\,r_{LH}}\right];$$
$R'_1$ takes the positive square root in the latter expression.
By inspection we can see that for the extreme cases $p=0$ and $p=1$,
$R'_1$ simplifies to $R_1$ for the one-track model.

Since the second term inside the radical is positive, we have that
$$R'_1>\frac{1}{2}\left[r_{HH}+r_{LL}+\sqrt{(r_{HH}-r_{LL})^2}\right]
  =\frac{1}{2}\left[r_{HH}+r_{LL}+|r_{HH}-r_{LL}|\right]
%  =\frac{1}{2}\left[2\,\max(r_{HH},r_{LL})\right]
  =\max(r_{HH},r_{LL}).$$
Since, for positive numbers $a$ and $b$, $\sqrt{a+b}<\sqrt{a}+\sqrt{b}$,
we also have that
$$R'_1<\frac{1}{2}\left[r_{HH}+r_{LL}+\sqrt{(r_{HH}-r_{LL})^2}
  +2\sqrt{r_{HL}\,r_{LH}}\right]
  =\max(r_{HH},r_{LL})+\sqrt{r_{HL}\,r_{LH}}.$$

\newpage

\section*{Tables}

\begin{table}[!h]
\begin{center}
\caption{Compartments and parameters of the two-track model}
\renewcommand{\baselinestretch}{1.75}
\small\normalsize
\begin{tabular}{|l|l|}
\hline
\multicolumn{2}{|c|}{Table of Compartments and Parameters}\\
\hline \hline
$H$ & high affinity susceptibles 
(i.e., voters highly susceptible to third party ideology)\\
\hline 
$L$ & low affinity susceptibles 
(i.e., voters barely susceptible to third party ideology)\\
\hline
$V_H$ &  third party voting individuals deriving from $H$\\
\hline
$V_L$ &  third party voter individuals deriving from $S$\\
\hline
$M$ & third party members (i.e., party officials, donors, volunteers)\\
\hline \hline
$p$ & proportion of the voting population $N$ entering $H$\\
\hline
$\beta_H$ & peer-driven recruitment rate of $H$ into $V_H$
by individuals in $V_H$, $V_L$ and $M$\\
\hline 
$\epsilon_H$ & linear recruitment rate of $V_H$ back into $H$ 
via secondary contacts\\
& (i.e., media and campaigning from opposing parties)\\
\hline
$\phi_H$ & recruitment rate of $V_H$ into $H$
by direct contact with individuals\\
& in the opposition classes (i.e., individuals in H and L)\\
\hline
$\beta_L$ & peer driven recruitment rate of $L$ into $V_L$ 
by individuals in $V_L$, $V_H$, and $M$\\ %(analogous to $\beta_H$)\\
\hline
$\epsilon_L$ & linear recruitment rate of $V_L$ back into $L$
via secondary contacts\\
& (i.e., media and campaigning from opposing parties)\\
\hline 
$\phi_L$ & recruitment rate of $V_L$ into $L$ by
direct contact with individuals\\
& in the opposition classes (i.e., individuals in $H$ and $L$)\\
\hline
$\alpha$ & factor by which the influence of party members $M$ in recruiting voters \\
& in $H$ and $L$ into $V_H$ and $V_L$ exceeds that of voters in $V_H$ and $V_L$ \\
\hline
%$\tau$ & factor by which the influence of members of the same affinity class\\
%& exceeds that of the other affinity class in discouraging third party voters
%$V_H$ and $V_L$\\
$\sigma$ & factor by which the influence of individuals upon members of a different\\
& affinity class is reduced in encouraging or discouraging third party voting\\
\hline
$\gamma$ & recruitment rate of $V_H$ and $V_L$ into $M$ by individuals in
$M$\\
\hline
$\mu$ & rate at which individuals enter or leave the voting system\\
\hline
\end{tabular}
\end{center}
\label{p2t}
\end{table}

\renewcommand{\baselinestretch}{2}
\small\normalsize

\newpage

\begin{table}[!h]
\begin{center}
\caption{Parameters of the one-track model} \label{p1t}
\begin{tabular}{|l|l|}
\hline $\beta$ & peer driven recruitment rate of $S$ into
$V$ by third party voters and members\\
\hline
$\epsilon$ & recruitment rate of $V$ back into $S$ via secondary contacts \\
& (i.e., media and campaigning from opposing parties)\\
\hline $\phi$ & recruitment rate of $V$ into $S$
by direct contact with susceptibles\\
\hline
$\alpha$&factor by which the recruitment rate of $S$ into $V$ by third party\\
& members exceeds the recruitment rate by individuals in $V$ \\
\hline $\gamma$ & recruitment rate of $V$ into $M$ by third party members\\
\hline $\mu$ & rate at which individuals enter or leave the voting system\\
\hline
\end{tabular}
\end{center}
\end{table}

\newpage

\begin{table}
%\renewcommand{\baselinestretch}{1.5}
%\small\normalsize
\begin{center}
\caption{Equilibria of one-track model}\label{equil1}
\begin{tabular}{|l|l|l|}
\hline
Equilibrium & Existence Cond. & Stability Cond. \\
\hline
\hline
Party-Free $E_1=(1,0,0)$ & always exists & $R_1<1$ \\
\hline
Member-Free $E_2=(\frac{1}{R_1},1-\frac{1}{R_1},0)$ & $R_1>1$ & $R_2<1$ \\
\hline
Survival $E_3=(1-\frac{\mu}{\gamma}-m^*_+,\frac{\mu}{\gamma},m^*_+)$ & 
(i) $R_2>1$, or & always stable \\
& (ii) $R_2<1$, $R_3>1$ & when it exists \\
\hline
Survival $E_4=(1-\frac{\mu}{\gamma}-m^*_-,\frac{\mu}{\gamma},m^*_-)$ &
$R_2<1$, $R_3>1$ & always unstable \\
\hline
\end{tabular}
\end{center}
% $R_3>1$ in the table above was originally $B<0$, $B^2-4C>0$
%\renewcommand{\baselinestretch}{1}
%\small\normalsize
\end{table}

\newpage

\begin{table}[!ht]
\begin{center}
\caption{Regions of Equilibrium Stability} \label{stabreg}
\begin{tabular}{|l|l|l|l|l|}
\hline
 & $E_1$ & $E_2$ & $E_3$ & $E_4$\\
\hline 
I   &  stable & unstable & does not exist & does not exist\\
\hline
II  & unstable & stable & does not exist & does not exist\\
\hline
III & unstable & unstable & stable & does not exist\\
\hline
IV  &  unstable & stable & stable & unstable\\
\hline
V   & stable & does not exist & stable & unstable\\
\hline
\end{tabular}
\end{center}
\end{table}

\newpage

\begin{table}
\centering
\caption{Estimated model parameters, for the seven U.S. states (and district)
used in the case study.  States are listed in decreasing order of proportional
membership $M$ to facilitate comparison with Figure~\ref{datafig}.  Initial
conditions $v(0)$ marked with asterisks *, as well as all model parameters
(except $\mu$), were estimated by data fitting as described in the main text.
All rates are given in units of 1/days; values below 0.01 are given in
scientific notation.}
\noindent\begin{tabular}{c|ccr|llll}
      & Initial & Final & Initial Conditions & & & & \\
$\!\!$State$\!\!$ &  time   & time  & $(S(0),V(0),M(0))$ & 
$\beta-\phi$ & $\alpha\beta$ & $\epsilon$ & $\gamma$ \\
\hline
ME & Jun 1998 & Nov 2006 & (91.12, 8.58, 0.31) &
5.27E--4 & 0.5022 & 1.00E--5 & 1.34E--4 \\
DC & Jun 2003 & Sep 2004 & (87.54, 8.61, 3.85) &
0.0181 & 1.9911 & 1.02 & 6.72E--4 \\
   & Jul 2005 & Aug 2006 & $\!\!$(79.52, 16.43*, 4.05) &
1.06E--3 & 0.200 & 0.468 & 0.0115 \\
   & Jan 2007 & Nov 2008 & (89.04, 6.63*, 4.33) &
1.66E--3 & 1.15 & 2.83 & 1.02E--7 \\
CA & Feb 1999 & Feb 2004 & (94.87, 3.09, 2.03) &
4.94E--4 & 0.7188 & 1.69E--5 & 3.73E--4 \\
   & Oct 2004 & May 2008 & (91.87, 4.80*, 3.32) &
0.0105 & 0.2193 & 0.162 & 1.94E--6 \\
OR & Jan 2001 & Oct 2004 & (88.45, 10.26, 1.29) &
1.56E--4 & 1.45 & 2.99E--5 & 1.29E--4 \\
   & Nov 2004 & Mar 2009 & (96.06, 1.70*, 2.23) &
0.0524 & 0.781 & 2.93 & 1.00E--6 \\
NY & Apr 1999 & Mar 2004 & (99.27, 0.73, 0.01) &
0.0119 & 0.352 & 4.00E--3 & 4.33E--4 \\
   & Nov 2004 & Mar 2008 & (98.65, 0.16*, 1.19) &
0.0567 & 0.283 & 3.54 & 1.00E--6 \\
MD & Aug 2000 & Mar 2009 & (98.17, 1.32, 0.01) &
0.0103 & 1.03 & 2.40E--3 & 1.93E--4 \\
PA & Apr 2001 & Nov 2006 & (97.33, 2.55, 0.11) &
1.01E--5 & 0.329 & 1.37E--6 & 9.85E--4 \\
\end{tabular}
\label{data1}
\end{table}

\newpage

\begin{table}
\centering
\caption{Reproductive numbers and stable equilibria for the one-track model,
calculated using parameter estimates from Table~\ref{data1}.  
Reproductive numbers not given are negative.
Equilibria are given as percentages of the target population.}
\begin{tabular}{cccccl}
State & Period & $R_1$ & $R_2$ & $R_3$ & Stable equilibria $(v^*,m^*)$ \\
\hline
ME & 1998--2006	& 9.44  & 2.62 & 43.5 &	$E_3=(34.15, \ 65.84)$ \\
DC & 2003--2004	& 0.18  & $-$  & 72.4 &	$E_1=(0,0)$, $E_3=(6.81, \ 89.30)$ \\
   & 2005--2006	& 0.0023& $-$  & 38.9 &	$E_1=(0,0)$, $E_3=(0.39, \ 98.63)$ \\
   & 2007--2008	&0.00059& $-$  & $-$  &	$E_1=(0, \ 0)$ \\
CA & 1999--2004	& 7.88  & 7.11 & 81.3 &	$E_3=(12.28, \ 87.71)$ \\
   & 2004--2008	& 0.065 & $-$  & $-$  &	$E_1=(0, \ 0)$ \\
OR & 2001--2004	& 2.06  & 1.45 & 71.8 &	$E_3=(35.40, \ 64.59)$ \\
   & 2004--2009	& 0.018 & $-$  & $-$  &	$E_1=(0, \ 0)$ \\
NY & 1999--2004	& 2.93  & 6.23 & 53.4 &	$E_3=(10.58, \ 89.27)$ \\
   & 2004--2008	& 0.016 & $-$  & $-$  &	$E_1=(0, \ 0)$ \\
MD & 2000--2009	& 4.20  & 3.21 & 72.1 &	$E_3=(23.76, \ 76.16)$ \\
PA & 2001--2006	& 0.21  & $-$  & 64.8 &	$E_1=(0,0)$, $E_3=(4.84,95.14)$	\\
\end{tabular}
\label{data2}
\end{table}

\end{document}